\documentclass[superscriptaddress]{revtex4-2} 
\usepackage{bm}
\usepackage{changes}
\usepackage{amsmath,amssymb}
\usepackage{graphicx}

\usepackage{titlesec}

\titleformat{\section}
{\raggedright\bfseries}
{\thesection}{1em}{}

\titleformat{\subsection}
{\raggedright\bfseries}
{\thesubsection}{1em}{}

\begin{document}

\preprint{APS/123-QED}

\title{Dissipative surface solitons in two-dimensional truncated lattices with linear gain and loss}

\author{Changming Huang}
\email{hcm123\_2004@126.com}
\affiliation{Department of Physics, Changzhi University, Changzhi, Shanxi 046011, China}

\author{Yan Wang}
\affiliation{Department of Physics, Changzhi University, Changzhi, Shanxi 046011, China}

\author{Pengcheng Liu}
\affiliation{Department of Physics, Changzhi University, Changzhi, Shanxi 046011, China}

\author{Qidong Fu}
\email{jellyfqd@sjtu.edu.cn}
\affiliation{School of Physics and Astronomy, Shanghai Jiao Tong University, Shanghai 200240, China}

\author{Liangwei Dong}
\email{dlw\_0@163.com}
\affiliation{Department of Physics, Zhejiang University of Science and Technology, Hangzhou, China, 310023}

\date{\today}

\date{\today}

\begin{abstract}
Dissipative solitons constitute a robust class of self-localized nonlinear states sustained by the dynamic balance between nonlinearity and gain-loss, possessing an intrinsic stability that stems from their fundamental attractor nature. When combined with lattice truncation, this balance gives rise to dissipative surface solitons (DSSs), whose existence and stability are jointly dictated by boundary-induced confinement and non-Hermitian dynamics. In two-dimensional truncated lattices with linear gain and loss, surface localization emerges within gap regimes, where families of DSSs bifurcate from linear surface localized gain modes as the nonlinearity increases. Increasing the number of waveguide rows at the interface enriches the diversity of supported surface modes in both linear and nonlinear regimes. Although multiple DSS families with distinct phase configurations may coexist within the same gap, their dynamical stability is strongly phase selective. These insights establish linear gain-loss engineering as a powerful mechanism for controlling nonlinear surface localization and provide practical guidelines for realizing robust nonlinear surface states in gain-loss-tailored photonic platforms.
\end{abstract}

\maketitle

\section{Introduction}
Dissipative solitons--self-organized, localized wave packets arising in non-conservative media--constitute a fundamental cornerstone in the study of non-equilibrium dynamics~\cite{belashov2005solitary,Purwins2010,Grelu2012,Peng2018,Rosanov2026}. Transcending the classical framework of conservative dynamics, these self-sustained entities are governed by a dual-balance architecture: while the interplay between nonlinearity and dispersion/diffraction ensures temporal/spatial localization, their stationarity is inextricably linked to a continuous energy exchange between localized gain and loss~\cite{Guo2016,malomed2022multidimensionalCSF}. This intricate stabilization regime, described by the complex Ginzburg-Landau equation~\cite{Aranson2002,ankiewicz2008dissipative,skarka2006stability}, has been rigorously studied across a broad range of physical systems. Such investigations span from ultrafast fiber lasers, where dissipative structures are mediated by saturable absorption~\cite{rosanov2002spatial,rosanov2005curvilinear,veretenov2016rotating, Veretenov2017}, to exotic configurations wherein linear gain is compensated by nonlinear loss~\cite{Hung2017,Perego2018}.

A fundamental tenet in this field is that the stability and existence of dissipative solitons are critically dictated by the underlying gain-loss landscape~\cite{Pernet2022,BlancoRedondo2023,Yan2025lpr}. It is a well-recognized constraint that homogeneous linear gain precludes the formation of stable one-dimensional ($1$D) solitons, as the unbounded growth of the background pedestal inevitably triggers a global instability~\cite{malomed1997stability}. Consequently, achieving robust localization necessitates the strategic implementation of spatially confined gain profiles. Within this localized paradigm, a wide range of nonlinear loss-mediated states has been elucidated. These range from $1$D defect modes and symmetry-breaking multi-peaked states~\cite{kartashov2010dissipativeOL,kartashov2011symmetryPRA,huang2020dissipativeOE} to higher-dimensional topological states, including vortex solitons and rotating clusters characterized by non-trivial winding numbers~\cite{kartashov2010vortexOL,borovkova2011rotatingOL,Huang2025CSF,veretenov2016rotatingPRL}.
Furthermore, the emergence of parity-time symmetry and non-Hermitian topology has redefined our understanding of dissipative phases. By converging classical nonlinear optics with quantum-inspired symmetries, these paradigms afford an unparalleled degree of structural protection against stochastic perturbations, thereby providing a fertile framework for the realization of exotic, self-organized solitonic states~\cite{musslimani2008opticalPRL,konotop2016nonlinearRMP,suchkovnonlinearLPR2016,kartashov2015topologicalPRL}.

Despite these advances, dissipative interfaces introduce additional constraints that remain less fully understood than their conservative counterparts. While surface solitons at the boundaries of optical lattices are well established in conservative nonlinear systems~\cite{kartashov2006surfacePRL,Smirnova2020,Ivanov2021,Tang2022}, their dissipative counterparts are governed by the combined requirements of gain-loss balance, nonlinear self-action, and dynamical stability. Previous studies on dissipative surface states~\cite{kartashov2010dissipativeEPL,he2012stableOL,huang2019dissipative,kartashov2019edgeOL} and localized gain structures~\cite{kartashov2011twoOL,borovkova2011rotatingOL,lobanov2010stableOL,kartashov2010vortexOL,li2024stablePRL,Huang2025CSF} have revealed a rich stability landscape, highlighting the need to clarify how boundary truncation organizes nonlinear localized states in gain-loss lattices.

In this work, we investigate geometry-controlled two-dimensional ($2$D) dissipative surface soliton (DSS) families in transversely truncated multi-row complex waveguide arrays with spatially distributed linear gain and loss. While dissipative surface solitons have been explored in several gain-loss systems~\cite{Mihalache2008PRE,zezyulin2023continuousOL,shen2025edgeFOP}, how the number of retained transverse waveguide rows controls their linear origin, global power balance, spatial localization and stability has not been systematically clarified.
By comparing single-, double- and triple-row truncated lattices, we show that transverse truncation acts as a geometric control parameter that enriches the accessible linear and nonlinear surface-state landscape. We show that DSSs bifurcate from gap-localized surface gain modes in the underlying linear spectrum. Importantly, our stability analysis and direct propagation simulations reveal that stable stationary propagation is restricted to in-phase DSSs, whereas out-of-phase and antisymmetric states become unstable. These results identify transverse truncation as a geometry-based route for controlling $2$D dissipative surface localization in non-Hermitian optical lattices.

\section{Model}
We consider the dynamical propagation of light beams along the $z$-axis in a cubic nonlinear medium featuring a truncated gain-loss lattice, described by the nonlinear Schr\"{o}dinger equation for the field amplitude $\Psi$:
\begin{equation}\label{eq1}
	i\frac{\partial \Psi}{\partial z} =-\frac{1}{2}\left(\frac{\partial^2\Psi}{\partial x^2}+\frac{\partial^2\Psi}{\partial y^2}\right)-(V_r-iV_i)\Psi-\sigma\left|\Psi\right|^2\Psi,
\end{equation}
Here, $x$ and $y$ are the transverse coordinates, while $z$ represents the longitudinal propagation direction. $\sigma$ is the Kerr nonlinearity coefficient, with $\sigma=1$ corresponding to focusing and $\sigma=-1$ to defocusing nonlinearity. $V_r-iV_i$ is the complex lattice potential, with $V_{r}$ representing the optical lattice profile and $V_{i}$ specifying the spatially distributed linear gain and loss. Their expressions are given by $V_{r}(x,y)=p_{r}\cos^2(x)\sum_{k=m}^{k=n}\exp\left[-(y-kd)^2/w^2\right]$ and $V_{i}(x,y)=\left[p_{i}\sin(2x)+\gamma\right]\sum_{k=m}^{k=n}\exp\left[-(y-kd)^2/w^2\right]$ for $x\geq 0$, and $V_{r,i}(x,y)=0$ for $x<0$. Here, $p_{r}$ denotes the depth of the two-dimensional optical lattice, $|p_{i}+\gamma|$ represents the amplitude of gain or loss, $w$ specifies the width of individual waveguides along the $y$ direction, and $d$ is the period of the waveguide array in the same direction. 
$k$ is an integer displacement index, so that the centre of the $k$-th waveguide row is located at $y=kd$. The integers $m$ and $n$ determine the lower and upper limits of the retained waveguide rows, respectively, and therefore specify the transverse truncation geometry.
We emphasize that, owing to the $\gamma$-dependent gain-loss landscape, the present complex lattice potential is neither globally nor partially $\mathcal{PT}$-symmetric, even before truncation.
Global $\mathcal{PT}$ symmetry requires $V(x,y)=V^*(-x,-y)$, whereas partial
$\mathcal{PT}$ symmetry requires $V(x,y)=V^{*}(-x,y)$. Thus, this truncated lattice is treated as a general non-Hermitian complex lattice with spatially distributed linear gain and loss.
The model is considered in the small-signal gain regime, where the gain and loss coefficients can be approximated as intensity independent. In this regime, gain saturation is a higher-order correction, while appreciable Kerr nonlinear phase shifts may still accumulate in highly nonlinear and tightly confined waveguide systems, such as AlGaAs, or doped active waveguide arrays~\cite{guo2009observation,dolgaleva2011compact,el2018nonNP}.

The proposed complex lattice potential can be realized in well-established photonic platforms. In femtosecond laser-written waveguide arrays, the refractive index profile is determined by the spatial arrangement and writing parameters of individual waveguides, enabling flexible control of discrete diffraction and wave propagation dynamics~\cite{streltsov2002study,szameit2010discrete}. The spatially distributed gain and loss can be implemented via optical pumping in doped dielectric waveguides or by controlled absorption engineering~\cite{ruter2010observation,belitsch2022gain}. Such waveguide lattice systems have been widely used for the experimental realization of complex non-Hermitian optical potentials~\cite{bandres2018topological,el2019dawn,ashida2020non}.

The dissipative surface solitons supported by Eq.~(\ref{eq1}) take the form $\Psi=\phi e^{i\beta z}=(\phi_r+i\phi_i)e^{i\beta z}$, where $\beta$ is the real-valued propagation constant, and $\phi$, $\phi_{r}$, and $\phi_{i}$ denote the complex field and its real and imaginary components, respectively. These components $\phi_{r}$ and $\phi_{i}$ are governed by the coupled equations:
\begin{align}
	\frac{1}{2}\left (\frac{\partial^2 \phi_{r}}{\partial x^2}+\frac{\partial^2 \phi_{r}}{\partial y^2}\right )+(V_{r}-\beta) \phi_{r}+V_{i} \phi_{i}+\sigma(\phi_{r}^{3}+\phi_{i}^{2} \phi_{r})=0, \label{eq2} \\
	\frac{1}{2}\left (\frac{\partial^2 \phi_{i}}{\partial x^2}+\frac{\partial^2 \phi_{i}}{\partial y^2}\right )+(V_{r}-\beta)\phi_{i}-V_{i} \phi_{r}+\sigma(\phi_{r}^{2} \phi_{i}+\phi_{i}^{3})=0. \label{eq3}
\end{align}
Eqs.~(\ref{eq2}) and (\ref{eq3}) are solved numerically via a Newton iteration scheme constrained by the condition
\begin{align}
	\frac{\partial U}{\partial z}=\iint2V_{i}(x,y)|\phi_r(x,y)+i\phi_i(x,y)|^2dxdy=0, \label{eq4}
\end{align}
ensuring convergence to dissipative soliton states maintained by the power balance, with
power $U=\iint[\phi^2_r(x,y)+\phi^2_i(x,y)]dxdy$.
In contrast to conventional Ginzburg-Landau-type dissipative solitons stabilized by nonlinear gain or loss, these soliton states discussed in this work are dissipative nonlinear localized modes supported by Kerr self-action, lattice confinement and the global power balance of spatially distributed linear gain and loss~\cite{he2012stableOL,zezyulin2011solitonsOL,wimmer2015observationNC}. In the following, we fix $p_r=4$, $p_i=1$, $d=4.5$, and $w=1.5$, and vary $k$, $\sigma$, and $\gamma$ to discuss the properties of the dissipative surface solitons.

The dynamical stability of dissipative surface solitons is examined via a linear stability analysis of the stationary solutions by introducing small perturbations of the form~\cite{kartashov2006surfacePRL,deconinck2006computing,yang2010nonlinear}
\begin{align}
	\Psi(x,y,z)=\{\phi_r(x,y)+i\phi_i(x,y)+[g_1(x,y)+ig_2(x,y)]e^{\tau z}\}e^{i\beta z},\label{eq5}
\end{align}
where $\tau$ is the perturbation growth rate (which can be complex), and $g_1(x,y)$ and $g_2(x,y)$ denote the real and imaginary components of the perturbation, respectively.
Substituting Eq. (\ref{eq5}) into Eq. (\ref{eq1}) and linearizing it around $\phi_r(x,y)$ and $\phi_i(x,y)$, one obtains a coupled set of linear eigenvalue equations:
\begin{align}
	\tau g_1 &= \left [-\frac{1}{2}\left (\frac{\partial^2}{\partial x^2}+\frac{\partial^2}{\partial y^2}\right ) - V_r + \beta - \sigma(3\phi_i^2 + \phi_r^2)\right ]g_2 -(2\sigma\phi_r\phi_i - V_i)g_1,  \label{eq6} \\
	\tau g_2 &= \left [+\frac{1}{2}\left (\frac{\partial^2}{\partial x^2}+\frac{\partial^2}{\partial y^2}\right ) + V_r - \beta + \sigma(3\phi_r^2 + \phi_i^2)\right ]g_1 +(2\sigma\phi_r\phi_i + V_i)g_2. \label{eq7}
\end{align}
The detailed derivation is given in Section \textcolor{black}{S1} of the Supporting Information, and the resulting eigenvalue problem is solved numerically using a finite-difference method. Linear stability requires $\text{Re}(\tau)=\tau_r\leq 0$ for all perturbation eigenvalues; any eigenvalue with $\tau_r>0$ renders the dissipative soliton unstable.

\section{Results and discussion}\label{sec3}
We first examine the linear spectra and eigenmode profiles of different transversely truncated complex lattices. Linear surface gain modes provide the seeds for the nonlinear dissipative surface soliton branches. We then consider single, dual and triple truncated arrays in sequence. For each geometry, we construct the corresponding DSS families and characterize their existence, stability, and propagation dynamics. This stepwise analysis reveals how the number of retained waveguide rows acts as a geometric control parameter for two-dimensional dissipative surface localization.

\subsection{Lattices and linear eigenvalues}
\begin{figure*}
	\centering
	\includegraphics[width=0.99\columnwidth]{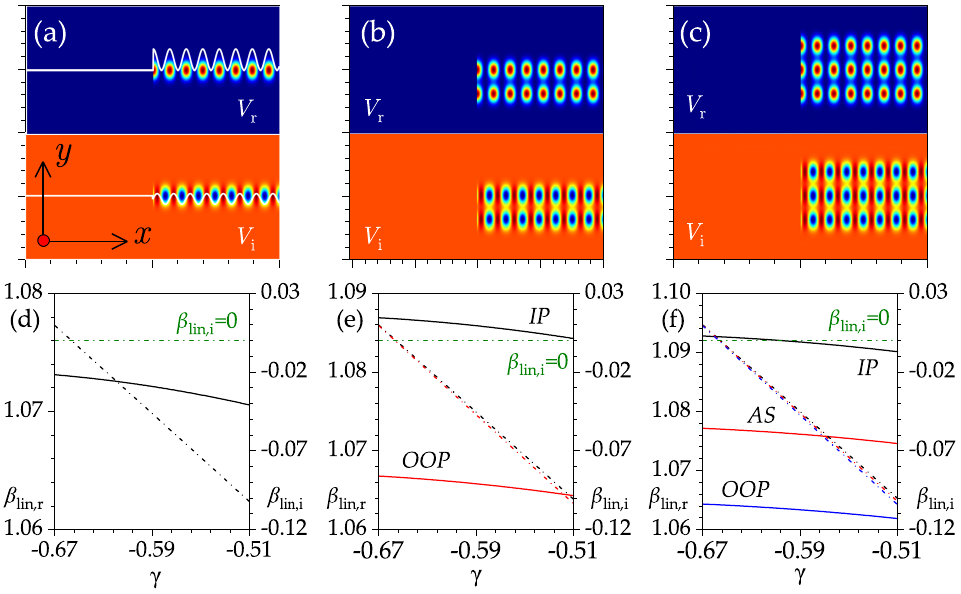}
	\caption{(a-c) Optical lattice profiles $V_r$ (top panels) and linear gain-loss landscapes $V_i$ (bottom panels) for transversely truncated single-, double-, and triple-waveguide arrays. White lines in (a) denote the $y=0$ cross-sectional profiles of $V_r$ and $V_i$.
		(d-f) Eigenvalue $\beta_\text{lin}=\beta_\text{lin,r}+i\beta_\text{lin,i}$ of linear localized surface modes supported at the lattice interface as functions of the parameter $\gamma$. Solid lines indicate the real parts of the eigenvalues ($\beta_\text{lin,r}$, left axis), while dashed lines denote the imaginary parts ($\beta_\text{lin,i}$, right axis); identical colors are used for the real and imaginary components of the same surface mode. In each panel of (a-c), $x\in[-24,+24]$, $y\in[-12,+12]$, $p_r=4$, $p_i=1$, $d=4.5$, $w=1.5$, and $\gamma=-0.6$.
		In (e) and (f), IP, OOP, and AS refer to in phase, out of phase, and antisymmetric, respectively.
	}\label{fig1}
\end{figure*}

In this work, we investigate the optical properties of dissipative surface solitons formed at an interface. Three types of transversely truncated waveguide arrays with their linear gain-loss profiles are considered and presented in Figs.~\ref{fig1}(a-c), showing that the structures are spatially truncated along both the $x$ and $y$ axes.
This truncated linear gain-loss landscape differs fundamentally from conventional $\mathcal{PT}$-symmetric or partially $\mathcal{PT}$-symmetric configurations \cite{musslimani2008opticalPRL,konotop2016nonlinearRMP,el2018nonNP,suchkovnonlinearLPR2016,Kang2023,Yan2025lpr}.

Before exploring the properties of the dissipative nonlinear surface states, it is crucial to first clarify the characteristics of their linear eigenvalues and the associated linear surface modes.
In the linear regime ($\sigma=0$), the proposed optical lattice supports multiple linear modes distributed across different waveguides, including a localized gain or loss surface mode at the truncated interface (see Section \textcolor{black}{S2}, Supporting Information).
The dependence of the surface-mode eigenvalues $\beta_\text{lin}=\beta_\text{lin,r}+i\beta_\text{lin,i}$ on the parameter $\gamma$ is presented in Figs.~\ref{fig1}(d-f).
As $\gamma$ increases, both the real ($\beta_\text{lin,r}$) and imaginary ($\beta_\text{lin,i}$) parts of the eigenvalue gradually decrease, with $\beta_\text{lin,i}$ changing sign from positive to negative.
This behavior indicates the existence of a critical value $\gamma_0$, such that $\beta_\text{lin,i}>0$ for $\gamma<\gamma_0$ and $\beta_\text{lin,i}<0$ for $\gamma>\gamma_0$, corresponding to linear loss and gain modes, respectively.
Notably, as the number of waveguide arrays along the $y$-axis increases, the interface supports a greater number of surface modes, accompanied by the emergence of new mode families.
A transversely truncated single-waveguide array supports a fundamental localized surface mode. Truncation of a two-waveguide array gives rise to both in-phase and out-of-phase surface modes, whereas truncation of a three-waveguide array additionally supports an antisymmetric mode (see Section \textcolor{black}{S2}, Supporting Information).
In such structures, the eigenvalue of the in-phase mode typically exceeds that of its out-of-phase counterpart [Figs.~\ref{fig1}(e-f)]. The eigenvalue of the newly emergent antisymmetric mode falls between these two, although the differences among all surface-mode eigenvalues are quantitatively small.

Near $\gamma_0=-0.657$, the linear surface-localized gain modes identified in Figs.~\ref{fig1}(d-f) and Fig.~\textcolor{black}{S1} provide the parent states for the nonlinear DSS families discussed below. The nonlinear branches are obtained by continuing these linear modes into the nonlinear regime, with the parent mode selected according to its spatial localization and phase symmetry. Thus, the fundamental DSS in the single truncated array bifurcates from the fundamental surface gain mode in Figs.~\ref{fig1}(d) and \textcolor{black}{S1}(a). The IP and OOP DSSs in the dual truncated arrays originate from the corresponding IP and OOP linear surface modes in Figs.~\ref{fig1}(e) and \textcolor{black}{S1}(c,d), whereas the IP, OOP and antisymmetric DSSs in the triple truncated arrays originate from the corresponding linear modes shown in Figs.~\ref{fig1}(f) and \textcolor{black}{S1}(e-g).

\begin{figure*}
	\centering
	\includegraphics[width=0.99\columnwidth]{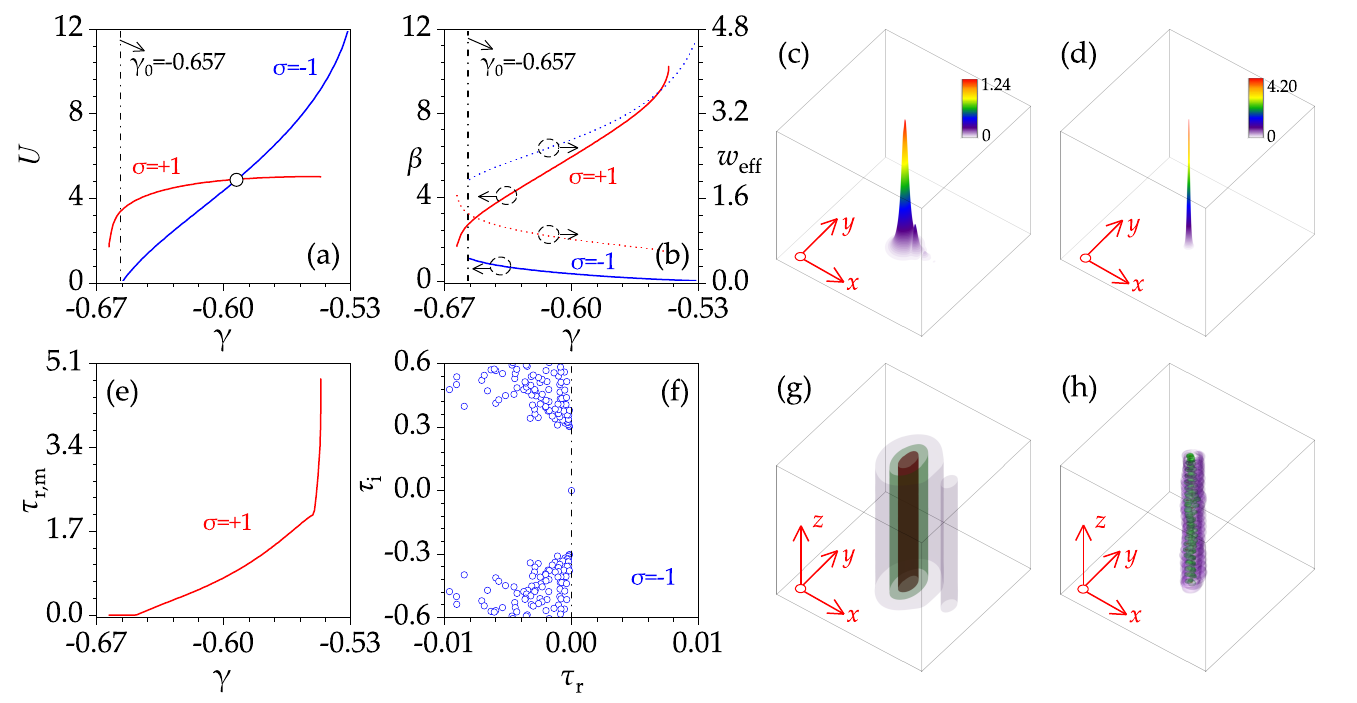}
	\caption{(a) Power $U$ and (b) propagation constants $\beta$ and effective width $w_\text{eff}$ versus $\gamma$ for dissipative surface solitons in a transversely truncated single-waveguide array. Dash-dotted lines in (a,b) mark $\gamma_0=-0.657$, where $\beta_i = 0$. Circles and arrows in (b) indicate which axis corresponds to each curve. (c,d) Representative field profiles $|\phi|$ of dissipative surface solitons, marked by circles in (a), for (c) $\gamma=-0.593$, $\sigma=-1$, and (d) $\gamma=-0.593$, $\sigma=+1$. (e) Maximum instability growth rate $\tau_\text{r,m}$ versus $\gamma$ for $\sigma=+1$. (f) Linear stability spectrum of a stable dissipative surface soliton at $\gamma=-0.593$ and $\sigma=-1$. (g,h) Isosurface visualizations of field modulus $|\phi|$ evolution along the propagation distance $z$, with $\gamma=-0.593$, $\sigma=-1$, $z\in[0,3000]$ in (g) and $\gamma=-0.593$, $\sigma=+1$, $z\in[0,1000]$ in (h). $x,y\in [-24,+24]$ in (c,d), and $x,y\in[-6,+6]$ in (g,h).}\label{fig2}
\end{figure*}
\begin{figure*}
	\centering
	\includegraphics[width=0.99\columnwidth]{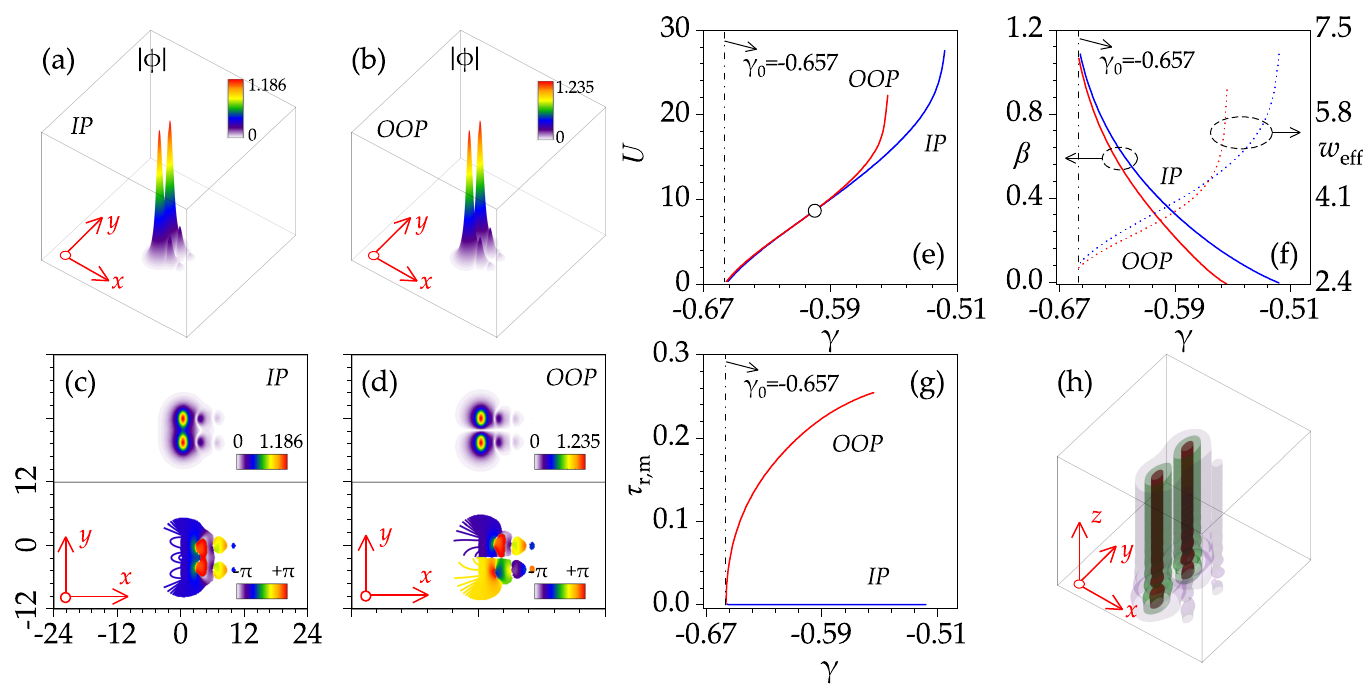}
	\caption{(a,b) Examples of field modulus profiles of in-phase (IP) and out-of-phase (OOP) dissipative surface solitons supported by a transversely truncated double-waveguide array. (c,d) Corresponding top-view field profiles (upper panels) and phase distributions (lower panels) are also shown. The IP state has the same phase sign at the two dominant lobes, whereas the OOP state has opposite phase signs, indicating a $\pi$-phase difference. Power $U$ (e), propagation constant $\beta$ and effective width $w_\text{eff}$ (f), and maximum instability growth rate $\tau_\text{r,m}$ (g) versus $\gamma$ for in-phase and out-of-phase solitons. Dash-dotted lines in (e-g) denote $\gamma$ values where $\beta_i = 0$. Circles and arrows in (f) indicate which axis corresponds to each curve. (h) Isosurface visualization of the field modulus evolution for an unstable out-of-phase soliton. $\gamma=-0.6$ in (a-d,h). $x,y\in[-24,+24]$ in (a,b), and $x,y\in [-8,+8]$ in (h). $\sigma=-1$ in all panels.}\label{fig3}
\end{figure*}
\begin{figure*}[h!]
	\centering
	\includegraphics[width=0.77\columnwidth]{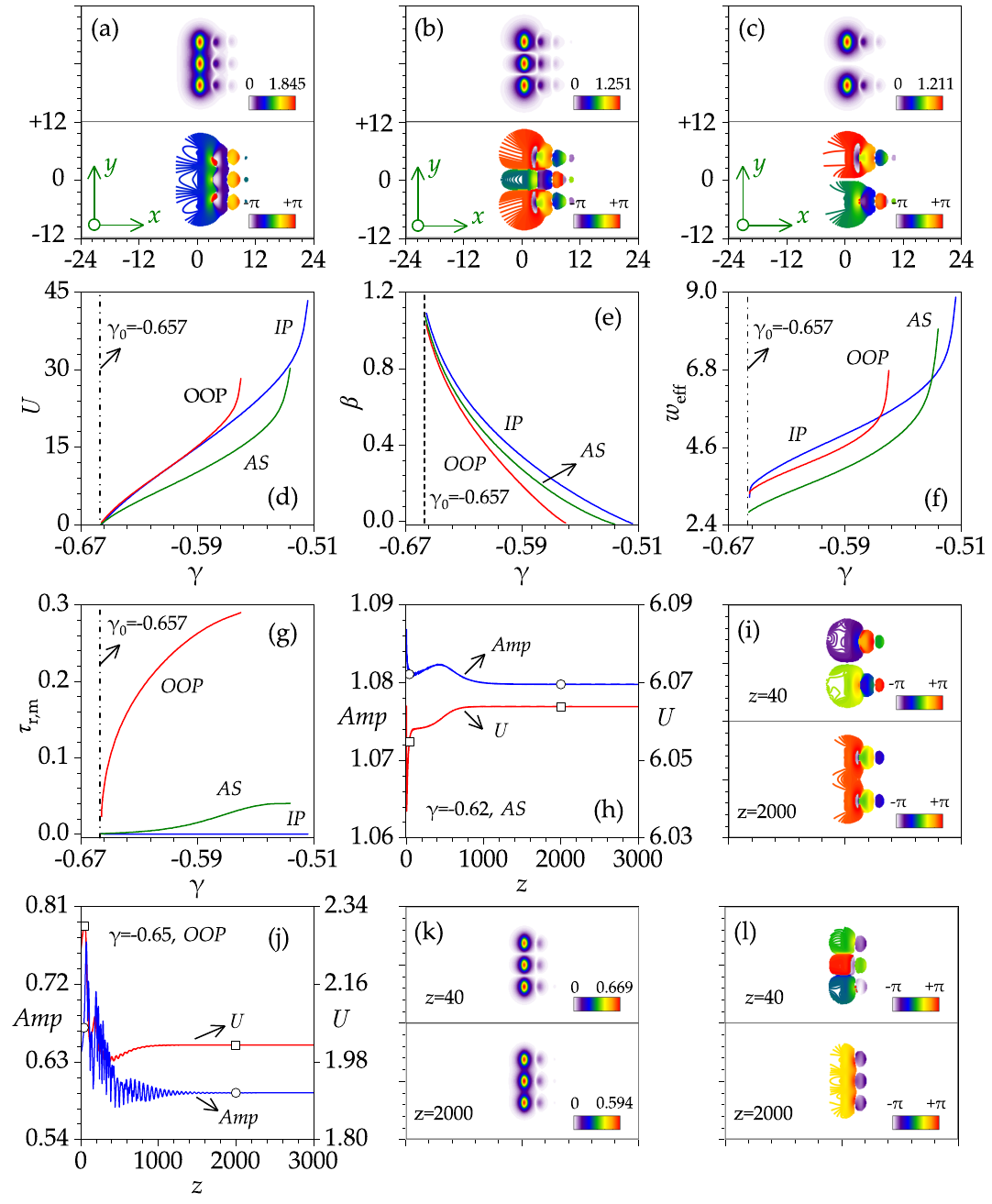}
	\caption{Representative field modulus profiles (upper panels) and corresponding phase distributions (lower panels) of dissipative surface solitons in triple truncated arrays: (a) in-phase DSSs, (b) out-of-phase DSSs, and (c) antisymmetric DSSs. For triple truncated arrays, the state is classified as in-phase when adjacent lobes have the same phase sign along both $y=+d/2$ and $y=-d/2$, and as out-of-phase when adjacent lobes have opposite phase signs. (d) Power $U$, (e) propagation constant $\beta$, (f) effective width $w_\text{eff}$, and (g) maximum instability growth rate $\tau_\text{r,m}$ as functions of the parameter $\gamma$ for the three DSS families. (h) Propagation dynamics of an unstable antisymmetric DSS, with (i) the corresponding phase distributions at selected propagation distances. (j) Propagation dynamics of an unstable out-of-phase DSS, with the corresponding field-modulus profiles in (k) and phase distributions in (l). $\gamma=-0.6$ in (a-c), $\gamma=-0.62$ in (h,i), $\gamma=-0.65$ in (j-l), $x\in[-24,+24]$ and $y\in[-12,+12]$ in (a-c,i,k,l). $\sigma=-1$ in all panels.} \label{fig4}
\end{figure*}

\subsection{DSSs in a single truncated array}
In a transversely truncated single-waveguide array, both focusing and defocusing nonlinearities are examined, with quantitative results--including the existence and stability characteristics--summarized in Fig.~\ref{fig2}.

Under focusing nonlinearity ($\sigma=+1$), the fundamental dissipative surface soliton exists only within a finite interval $[\gamma_{lc},\gamma_{rc}]$.
Both the power $U$ and the propagation constant $\beta$ increase monotonically with $\gamma$ [Figs.~\ref{fig2}(a,b)]. As $\gamma$ approaches its lower cutoff $\gamma_{lc}\approx-0.663$, the slope of $U(\gamma)$ becomes nearly vertical, whereas near the upper cutoff $\gamma_{rc}\approx-0.547$ the slope of $\beta(\gamma)$ similarly steepens. These behaviors collectively indicate the emergence of two sharply defined existence boundaries [Figs.~\ref{fig2}(a,b)].
Notably, these dissipative surface solitons arise purely from nonlinear effects, as indicated by the fact that their lower cutoff $\gamma_{lc}<\gamma_0$ and that their existence requires a finite power threshold.

For defocusing nonlinearity ($\sigma=-1$), fundamental dissipative surface solitons likewise exist only within a range of $\gamma$, yet their existence properties differ fundamentally from those of solitons in the focusing regime [also see Figs.~\ref{fig2}(a,b)].
The power $U$ of dissipative solitons increases monotonically with $\gamma$, whereas its propagation constant $\beta$ decreases monotonically as $\gamma$ increases.
As $\gamma$ approaches the upper cutoff $\gamma_{rc}$, the soliton's propagation constant asymptotically tends to zero. The lower cutoff $\gamma_{lc}$ occurs near $\gamma_0$, and as $\gamma$ approaches this value, the power $U$ of fundamental dissipative surface soliton diminishes to zero. This characteristic indicates a bifurcation of the dissipative surface soliton from the linear gain surface mode.

Figures~\ref{fig2}(c,d) show the typical field modulus of fundamental dissipative surface solitons in the transversely truncated single‐waveguide array for $\sigma=-1$ and $\sigma=+1$, respectively. It can be observed that, for the same $\gamma$, the surface soliton in the focusing nonlinear regime is significantly more localized than its defocusing counterpart.
The localization of the soliton is quantified by the inverse participation ratio $P\equiv U^{-2}\cdot\left[\iint |\phi(x,y)|^4dxdy \right]$, and the effective width is given by $w_\text{eff}=P^{-1/2}$.
As seen in Fig.~\ref{fig2}(b), for $\sigma = +1$, the soliton becomes more localized with increasing $\gamma$, whereas for $\sigma = -1$, the soliton broadens as $\gamma$ increases.

The instability growth rate of the solitons mentioned above is obtained by solving the linear eigenvalue equations (\ref{eq6}) and (\ref{eq7}). For $\sigma=+1$, the dissipative surface solitons are predominantly unstable, except in a narrow region near the lower critical value, as shown in Fig.~\ref{fig2}(e). In contrast, for $\sigma = -1$, fundamental dissipative surface solitons remain stable in their entire existence domain. The spectrum of a representative stable dissipative soliton is shown in Fig.~\ref{fig2}(f), where the real part of the maximum instability growth rate, $\tau_{r,m}$, is less than or equal to 0.

To validate the linear stability analys of the fundamental dissipative solitons, we performed propagation simulations for all solitons using an input beam $\Psi(x,y,z=0)=\phi(x,y)[1+\varrho(x,y)]$, where $\phi(x,y)$ denotes the stationary fundamental soliton and $\varrho(x,y)$ represents an added two-dimensional random perturbation.
The dynamical evolution of the dissipative solitons is fully consistent with the predictions of the linear stability analysis. Specifically, stable fundamental dissipative surface solitons preserve their profiles over long propagation distances, whereas unstable ones become distorted after only a short distance [Figs.~\ref{fig2}(g,h), also see Section \textcolor{black}{S3}, Supporting Information].

\subsection{DSSs in dual truncated arrays}
As the waveguide array extends in the $y$-direction, the family of dissipative surface solitons supported at the interface correspondingly expands. For displacement indices $k \in \{-1,0\}$, the lattice forms a transversely truncated dual-waveguide array. In this optical lattice, both in-phase (IP) and out-of-phase (OOP) dissipative surface solitons emerge along the $y$-direction [Figs.~\ref{fig3}(a-d)].  The field-modulus profiles show that the in-phase dissipative surface soliton retains a finite amplitude at $y=-d/2$, in contrast to the out-of-phase soliton, which vanishes at this point. The phase distributions [$\varphi=atan(\phi_i/\phi_r)$] reinforce this classification: matching phase peaks on either side of $y=-d/2$ correspond to an in-phase soliton, whereas opposite signs mark the out-of-phase state. These states can be viewed as composites of two fundamental dissipative surface solitons carrying different phases. In what follows, we focus on the defocusing regime ($\sigma=-1$), where stable DSS families extend over broader parameter ranges and their stability properties can be resolved more clearly. We have also examined the focusing case ($\sigma=+1$) for the double- and triple-row truncated lattices. Similar to the single-row array, only the in-phase DSSs display a very narrow stability window close to the left edge of their existence domain, whereas the other focusing DSSs are predominantly unstable. Moreover, these focusing states are also purely nonlinear modes, rather than nonlinear continuations of linear surface-localized gain or loss modes.

The existence and fundamental properties of the in-phase and out-of-phase dissipative surface solitons for $\sigma=-1$ are shown in Figs.~\ref{fig3}(e,f), revealing critical insights into the system's nonlinear dynamics.
The power $U$ of dissipative surface soliton are plotted as a function of the continuous control parameter $\gamma$ [Fig.~\ref{fig3}(e)]. Both IP (blue line) and OOP (red line) solitons demonstrate a monotonic increase in $U$ as $\gamma$ approaches less negative values (i.e., moving towards $\gamma \approx -0.51$).
As $\gamma$ asymptotically approaches $\gamma_0=-0.657$, the power $U$ for both soliton families diminishes to zero, revealing their bifurcation from the associated linear gain modes. Near this point, the IP surface solitons exhibit slightly higher power than their OOP counterparts, a trend that mirrors the linear limit in which the eigenvalue of the IP surface mode lies slightly above that of the OOP mode. Moreover, the upper cutoff of the OOP surface solitons ($\gamma_{rc}\approx-0.554$) is lower than that of the IP solitons ($\gamma_{rc}\approx-0.518$), and both families exhibit a pronounced increase in power as they approach their respective upper cutoffs.
Both classes of localized dissipative surface solitons are strictly confined to the regime $\gamma_0<\gamma \leq \gamma_{rc}$.
Within this bounded interval, the control parameter $\gamma$ sets the appropriate nonlinear strength needed to counteract dissipation and sustain a self-localized wave. We also observe that the propagation constants $\beta$ of both types of solitons decrease monotonically as $\gamma$ increases, accompanied by a monotonic broadening of their effective widths $w_\text{eff}$ [Fig.~\ref{fig3}(f)]. In the entire parameter range, the propagation constant $\beta$ of the OOP surface soliton remains lower than that of the IP soliton, and the difference grows steadily with increasing nonlinear strength. Except in the vicinity of the upper cutoff, the IP surface soliton generally exhibits a larger width than the OOP soliton.

While the fundamental dissipative surface solitons remain entirely stable for $\sigma=-1$, their composite states with different phase relations display pronounced stability contrasts. The linear stability analysis indicates that OOP surface solitons are unstable in their existence domain, whereas IP surface solitons preserve stability over their full range [Fig.~\ref{fig3}(g)].
This behavior can be physically understood by noting that the stability contrast arises because the nearly uniform phase of in-phase states enables the defocusing nonlinearity to generate balanced transverse forces along the $y$ direction. By contrast, out-of-phase states feature an almost $\pi$ phase jump along $y$ direction, which introduces a strong local phase gradient and produces pronounced positive and negative power-flow peaks in the narrow region at $y=-d/2$. This sharp phase discontinuity produces a pronounced peak in the transverse power flow near the node, leading to strongly unbalanced repulsive forces and, consequently, intrinsic instability (see Section \textcolor{black}{S4}, Supporting Information).

Direct propagation simulations for dual truncated arrays show that, in the defocusing regime ($\sigma=-1$), in-phase DSSs are dynamically robust, with their amplitude, power and localized profile remaining nearly unchanged up to long propagation distances. By contrast, out-of-phase DSSs undergo a transient relaxation in which the initial phase jump disappears, indicating their instability and eventual evolution into stable in-phase DSSs [Fig.~\ref{fig3}(h), also see Section \textcolor{black}{S5}, Supporting Information].

\subsection{DSSs in triple truncated arrays}
We also investigate the existence and dynamical stability of dissipative surface solitons supported by a triple truncated waveguide array with $k\in \{-1,0,+1\}$. Representative stationary field modulus distributions for three different phase configurations: the in-phase (IP) state, the out-of-phase (OOP) state, and antisymmetric (AS) state, are shown in Figs.~\ref{fig4}(a-c), respectively. In each panel, the upper row shows the field modulus $|\phi|$, while the lower row depicts the corresponding phase $\varphi$. These surface solitons are strongly localized near the truncated edge, with distinct internal phase structures.

The power $U$, propagation constant $\beta$, and effective width $w_\text{eff}$ of the three types of dissipative surface solitons are depicted as functions of the parameter $\gamma$ in Figs.~\ref{fig4}(d-f), respectively.
All three soliton families bifurcate from the linear limit at $\gamma \approx -0.657$, indicated by the vertical dashed line, and persist within overlapping but distinct existence domains.
The existence domains decrease successively from the in-phase soliton to the antisymmetric soliton and further to the out-of-phase soliton. Near the lower cutoff, the powers of the in-phase and out-of-phase surface solitons are comparable and are approximately $1.5$ times larger than that of the antisymmetric surface soliton [Fig.~\ref{fig4}(d)]. In the linear regime, the eigenvalue of the in-phase mode exceeds that of the antisymmetric mode, which in turn is higher than that of the out-of-phase mode. This hierarchy is preserved in the nonlinear regime, where the corresponding propagation constants follow the same ordering [Fig.~\ref{fig4}(e)].
Except in the vicinity of the upper cutoff, the in-phase dissipative solitons exhibit the largest width, while the antisymmetric solitons possess the smallest one. As the upper cutoff is approached, the slope of the $w_\text{eff}(\gamma)$ curve increases rapidly and tends toward a nearly vertical orientation [Fig.~\ref{fig4}(f)].

Within this waveguide configuration, we likewise focus on the stability of the supported surface dissipative solitons. Linear stability analysis reveals that the in-phase surface dissipative solitons remain stable in their entire existence domain, whereas both the out-of-phase and antisymmetric surface solitons are unstable over the full parameter range [Fig.~\ref{fig4}(g)].
Direct propagation simulations show that unstable antisymmetric and out-of-phase surface solitons do not preserve their initial phase configurations, but typically evolve toward in-phase localized states during propagation [Figs~\ref{fig4}(h-l)].
The underlying mechanism governing their dynamical stability closely resembles that of dissipative solitons in dual truncated arrays, and it further suggests that introducing a transverse phase translation along the $y$ direction in engineered waveguide arrays may enable the realization of stable out-of-phase solitons.

Finally, to complement the stability analysis across all truncated geometries, we performed additional propagation simulations, as presented in Sections \textcolor{black}{S6} and \textcolor{black}{S7} of the Supporting Information. These results clarify the instability dynamics near the stability boundaries in different truncated waveguide arrays and demonstrate that stable dissipative surface solitons can be dynamically excited from weak Gaussian inputs close to the zero-intensity background.

\section{Conclusions}\label{sec5}
In conclusion, we have elucidated the formation, bifurcation, and stability properties of dissipative surface solitons in truncated two-dimensional waveguide lattices with linear gain and loss. Surface-localized linear gain modes emerge within spectral gaps due to lattice truncation, and families of dissipative surface solitons are shown to bifurcate from these modes as nonlinearity is introduced, establishing a clear spectral origin of nonlinear surface localization in non-Hermitian lattices.
The stability of dissipative surface solitons is found to be strongly influenced by both the type of nonlinearity and the phase configuration. While focusing nonlinearity generally destabilizes fundamental surface solitons, defocusing nonlinearity robustly supports stable dissipative surface localization over broad parameter ranges. In multi-row truncated lattices, although multiple families of dissipative surface solitons with different phase relations may coexist, their dynamical stability exhibits pronounced phase selectivity: only in-phase surface solitons remain stable, whereas out-of-phase and antisymmetric states undergo spontaneous symmetry restoration and evolve into stable in-phase configurations.

\section*{Supporting Information}
Supporting Information is available from the Online.

\section*{Acknowledgements}
This work was supported by the Applied Basic Research Program of Shanxi Province (202303021211191, 202303021212274), National Natural Science Foundation of China (62575264, 12404385), and China Postdoctoral Science Foundation (BX20230218, 2024M751950).

\appendix
\nocite{*}
\bibliographystyle{elsarticle-num}

\begin{thebibliography}{10}
	\expandafter\ifx\csname url\endcsname\relax
	\def\url#1{\texttt{#1}}\fi
	\expandafter\ifx\csname urlprefix\endcsname\relax\def\urlprefix{URL }\fi
	\expandafter\ifx\csname href\endcsname\relax
	\def\href#1#2{#2} \def\path#1{#1}\fi
	
	\bibitem{belashov2005solitary}
	V.~Y. Belashov, S.~V. Vladimirov, Solitary Waves in Dispersive Complex Media:
	Theory{\textperiodcentered} Simulation{\textperiodcentered} Applications,
	Springer Berlin Heidelberg, 2005.
	
	\bibitem{Purwins2010}
	H.-G. Purwins, H.~B\"{o}deker, S.~Amiranashvili, Dissipative solitons, Adv.
	Phys. 59~(5) (2010) 485–701.
	
	\bibitem{Grelu2012}
	P.~Grelu, N.~Akhmediev, Dissipative solitons for mode-locked lasers, Nat.
	Photonics 6~(2) (2012) 84–92.
	
	\bibitem{Peng2018}
	J.~Peng, H.~Zeng, Build-up of dissipative optical soliton molecules via diverse
	soliton interactions, Laser Photonics Rev. 12~(8) (2018) 1800009.
	
	\bibitem{Rosanov2026}
	N.~N. Rosanov, S.~V. Fedorov, N.~A. Veretenov, Dissipative optical solitons:
	From scalar 1d- to 2d- and 3d-, topological and vector solitons, Phys. Rep.
	1162 (2026) 1–60.
	
	\bibitem{Guo2016}
	H.~Guo, M.~Karpov, E.~Lucas, A.~Kordts, M.~Pfeiffer, V.~Brasch, G.~Lihachev,
	V.~Lobanov, M.~Gorodetsky, T.~Kippenberg, Universal dynamics and
	deterministic switching of dissipative kerr solitons in optical
	microresonators, Nat. Phys. 13~(1) (2016) 94–102.
	
	\bibitem{malomed2022multidimensionalCSF}
	B.~A. Malomed, Multidimensional dissipative solitons and solitary vortices,
	Chaos, Solitons Fractals 163 (2022) 112526.
	
	\bibitem{Aranson2002}
	I.~S. Aranson, L.~Kramer, The world of the complex ginzburg-landau equation,
	Rev. Mod. Phys. 74~(1) (2002) 99–143.
	
	\bibitem{ankiewicz2008dissipative}
	A.~Ankiewicz, N.~Akhmediev, Dissipative solitons: from optics to biology and
	medicine, Springer, 2008.
	
	\bibitem{skarka2006stability}
	V.~Skarka, N.~Aleksi{\'c}, Stability criterion for dissipative soliton
	solutions of the one-, two-, and three-dimensional complex cubic-quintic
	ginzburg-landau equations, Phys. Rev. Lett. 96~(1) (2006) 013903.
	
	\bibitem{rosanov2002spatial}
	N.~N. Rosanov, Spatial hysteresis and optical patterns, Springer Science \&
	Business Media, 2002.
	
	\bibitem{rosanov2005curvilinear}
	N.~N. Rosanov, S.~V. Fedorov, A.~N. Shatsev, Curvilinear motion of multivortex
	laser-soliton complexes with strong and weak coupling, Phys. Rev. Lett.
	95~(5) (2005) 053903.
	
	\bibitem{veretenov2016rotating}
	N.~Veretenov, N.~Rosanov, S.~Fedorov, Rotating and precessing
	dissipative-optical-topological-3d solitons, Phys. Rev. Lett. 117~(18) (2016)
	183901.
	
	\bibitem{Veretenov2017}
	N.~Veretenov, S.~Fedorov, N.~Rosanov, Topological vortex and knotted
	dissipative optical 3d solitons generated by 2d vortex solitons, Phys. Rev.
	Lett. 119~(26) (2017) 263901.
	
	\bibitem{Hung2017}
	N.~V. Hung, K.~Zegadlo, A.~Ramaniuk, V.~V. Konotop, M.~Trippenbach,
	Modulational instability of coupled ring waveguides with linear gain and
	nonlinear loss, Sci. Rep. 7~(1) (2017) 4089.
	
	\bibitem{Perego2018}
	A.~M. Perego, S.~K. Turitsyn, K.~Staliunas, Gain through losses in nonlinear
	optics, Light: Sci. Appl. 7~(1) (2018) 43.
	
	\bibitem{Pernet2022}
	N.~Pernet, P.~St-Jean, D.~D. Solnyshkov, G.~Malpuech, N.~Carlon~Zambon,
	Q.~Fontaine, B.~Real, O.~Jamadi, A.~Lemaître, M.~Morassi, L.~Le~Gratiet,
	T.~Baptiste, A.~Harouri, I.~Sagnes, A.~Amo, S.~Ravets, J.~Bloch, Gap solitons
	in a one-dimensional driven-dissipative topological lattice, Nat. Phys.
	18~(6) (2022) 678–684.
	
	\bibitem{BlancoRedondo2023}
	A.~Blanco-Redondo, C.~M. de~Sterke, C.~Xu, S.~Wabnitz, S.~K. Turitsyn, The
	bright prospects of optical solitons after 50 years, Nat. Photonics 17~(11)
	(2023) 937–942.
	
	\bibitem{Yan2025lpr}
	W.~Yan, R.~Chen, W.~Liu, Y.~Tan, Y.~Kivshar, F.~Chen, Tunable edge states and
	topological solitons in non-hermitian photonic lattices, Laser Photonics Rev.
	19~(15) (2025) 2402126.
	
	\bibitem{malomed1997stability}
	B.~A. Malomed, M.~G{\"o}lles, I.~M. Uzunov, F.~Lederer, Stability and
	interactions of pulses in simplified ginzburg-landau equations, Phys. Scr.
	55~(1) (1997) 73.
	
	\bibitem{kartashov2010dissipativeOL}
	Y.~V. Kartashov, V.~V. Konotop, V.~A. Vysloukh, L.~Torner, Dissipative defect
	modes in periodic structures, Opt. Lett. 35~(10) (2010) 1638--1640.
	
	\bibitem{kartashov2011symmetryPRA}
	Y.~V. Kartashov, V.~V. Konotop, V.~A. Vysloukh, Symmetry breaking and
	multipeaked solitons in inhomogeneous gain landscapes, Phys. Rev. A 83~(4)
	(2011) 041806.
	
	\bibitem{huang2020dissipativeOE}
	C.~Huang, C.~Li, L.~Dong, Dissipative solitons supported by transversal
	single-or three-channel amplifying chirped lattices, Opt. Express 28~(14)
	(2020) 21134--21142.
	
	\bibitem{kartashov2010vortexOL}
	Y.~V. Kartashov, V.~V. Konotop, V.~A. Vysloukh, L.~Torner, Vortex lattice
	solitons supported by localized gain, Opt. Lett. 35~(19) (2010) 3177--3179.
	
	\bibitem{borovkova2011rotatingOL}
	O.~V. Borovkova, V.~E. Lobanov, Y.~V. Kartashov, L.~Torner, Rotating vortex
	solitons supported by localized gain, Opt. Lett. 36~(10) (2011) 1936--1938.
	
	\bibitem{Huang2025CSF}
	C.~Huang, Q.~Fu, L.~Ma, Stable high-charge vortex dissipative solitons in
	azimuthally modulated waveguide arrays with localized gain, Chaos, Solitons
	Fractals 201 (2025) 117314.
	
	\bibitem{veretenov2016rotatingPRL}
	N.~Veretenov, N.~Rosanov, S.~Fedorov, Rotating and precessing
	dissipative-optical-topological-3d solitons, Phys. Rev. Lett. 117~(18) (2016)
	183901.
	
	\bibitem{musslimani2008opticalPRL}
	Z.~H. Musslimani, K.~G. Makris, R.~El-Ganainy, D.~N. Christodoulides, Optical
	solitons in \text{PT} periodic potentials, Phys. Rev. Lett. 100~(3) (2008)
	030402.
	
	\bibitem{konotop2016nonlinearRMP}
	V.~V. Konotop, J.~Yang, D.~A. Zezyulin, Nonlinear waves in \text{PT}-symmetric
	systems, Rev. Mod. Phys. 88~(3) (2016) 035002.
	
	\bibitem{suchkovnonlinearLPR2016}
	S.~V. Suchkov, A.~A. Sukhorukov, J.~Huang, S.~V. Dmitriev, C.~Lee, Y.~S.
	Kivshar, Nonlinear switching and solitons in \text{PT}-symmetric photonic
	systems, Laser Photonics Rev. 10~(2) (2016) 177--213.
	
	\bibitem{kartashov2015topologicalPRL}
	Y.~V. Kartashov, V.~V. Konotop, L.~Torner, Topological states in
	partially-\text{PT}-symmetric azimuthal potentials, Phys. Rev. Lett. 115~(19)
	(2015) 193902.
	
	\bibitem{kartashov2006surfacePRL}
	Y.~V. Kartashov, V.~A. Vysloukh, L.~Torner, Surface gap solitons, Phys. Rev.
	Lett. 96~(7) (2006) 073901.
	
	\bibitem{Smirnova2020}
	D.~Smirnova, D.~Leykam, Y.~Chong, Y.~Kivshar, Nonlinear topological photonics,
	Appl. Phys. Rev. 7~(2) (2020).
	
	\bibitem{Ivanov2021}
	S.~K. Ivanov, Y.~V. Kartashov, A.~Szameit, L.~Torner, V.~V. Konotop, Floquet
	edge multicolor solitons, Laser Photonics Rev. 16~(3) (2022) 2100398.
	
	\bibitem{Tang2022}
	G.-J. Tang, X.-T. He, F.-L. Shi, J.-W. Liu, X.-D. Chen, J.-W. Dong, Topological
	photonic crystals: physics, designs, and applications, Laser Photonics Rev.
	16~(4) (2022) 2100300.
	
	\bibitem{kartashov2010dissipativeEPL}
	Y.~V. Kartashov, V.~V. Konotop, V.~A. Vysloukh, Dissipative surface solitons in
	periodic structures, Europhys. Lett. 91~(3) (2010) 34003.
	
	\bibitem{he2012stableOL}
	Y.~He, D.~Mihalache, X.~Zhu, L.~Guo, Y.~V. Kartashov, Stable surface solitons
	in truncated complex potentials, Opt. Lett. 37~(13) (2012) 2526--2528.
	
	\bibitem{huang2019dissipative}
	C.~Huang, L.~Dong, Dissipative surface solitons in a nonlinear fractional
	schr{\"o}dinger equation, Opt. Lett. 44~(22) (2019) 5438--5441.
	
	\bibitem{kartashov2019edgeOL}
	Y.~V. Kartashov, V.~A. Vysloukh, Edge and bulk dissipative solitons in
	modulated \text{PT}-symmetric waveguide arrays, Opt. Lett. 44~(4) (2019)
	791--794.
	
	\bibitem{kartashov2011twoOL}
	Y.~V. Kartashov, V.~V. Konotop, V.~A. Vysloukh, Two-dimensional dissipative
	solitons supported by localized gain, Opt. Lett. 36~(1) (2011) 82--84.
	
	\bibitem{lobanov2010stableOL}
	V.~E. Lobanov, Y.~V. Kartashov, V.~A. Vysloukh, L.~Torner, Stable radially
	symmetric and azimuthally modulated vortex solitons supported by localized
	gain, Opt. Lett. 36~(1) (2010) 85--87.
	
	\bibitem{li2024stablePRL}
	C.~Li, Y.~V. Kartashov, Stable vortex solitons sustained by localized gain in a
	cubic medium, Phys. Rev. Lett. 132~(21) (2024) 213802.
	
	\bibitem{Mihalache2008PRE}
	D.~Mihalache, D.~Mazilu, F.~Lederer, Y.~S. Kivshar, Spatiotemporal dissipative
	solitons in two-dimensional photonic lattices, Phys. Rev. E 78~(5) (Nov.
	2008).
	
	\bibitem{zezyulin2023continuousOL}
	D.~A. Zezyulin, Continuous families of non-hermitian surface solitons, Opt.
	Lett. 48~(18) (2023) 4773--4776.
	
	\bibitem{shen2025edgeFOP}
	S.~Shen, M.~R. Beli{\'c}, Y.~Zhang, Y.~Li, T.~Wang, Z.-N. Tian, Q.-D. Chen,
	Edge solitons in non-hermitian photonic lattices with type-ii dirac cones,
	Front. Phys. 20~(4) (2025) 042203.
	
	\bibitem{guo2009observation}
	A.~Guo, G.~J. Salamo, D.~Duchesne, R.~Morandotti, M.~Volatier-Ravat, V.~Aimez,
	G.~A. Siviloglou, D.~N. Christodoulides, Observation of \text{PT}-symmetry
	breaking in complex optical potentials, Phys. Rev. Lett 103~(9) (2009)
	093902.
	
	\bibitem{dolgaleva2011compact}
	K.~Dolgaleva, W.~C. Ng, L.~Qian, J.~S. Aitchison, Compact highly-nonlinear
	\text{AlGaAs} waveguides for efficient wavelength conversion, Opt. Express
	19~(13) (2011) 12440--12455.
	
	\bibitem{el2018nonNP}
	R.~El-Ganainy, K.~G. Makris, M.~Khajavikhan, Z.~H. Musslimani, S.~Rotter, D.~N.
	Christodoulides, Non-hermitian physics and \text{PT} symmetry, Nat. Phys.
	14~(1) (2018) 11--19.
	
	\bibitem{streltsov2002study}
	A.~M. Streltsov, N.~F. Borrelli, Study of femtosecond-laser-written waveguides
	in glasses, J. Opt. Soc. Am. B 19~(10) (2002) 2496--2504.
	
	\bibitem{szameit2010discrete}
	A.~Szameit, S.~Nolte, Discrete optics in femtosecond-laser-written photonic
	structures, J. Phys. B:At., Mol. Opt. Phys. 43~(16) (2010) 163001.
	
	\bibitem{ruter2010observation}
	C.~E. R{\"u}ter, K.~G. Makris, R.~El-Ganainy, D.~N. Christodoulides, M.~Segev,
	D.~Kip, Observation of parity--time symmetry in optics, Nat. Phys. 6~(3)
	(2010) 192--195.
	
	\bibitem{belitsch2022gain}
	M.~Belitsch, D.~N. Dirin, M.~V. Kovalenko, K.~Pichler, S.~Rotter, A.~Ghalgaoui,
	H.~Ditlbacher, A.~Hohenau, J.~R. Krenn, Gain and lasing from cdse\/cds
	nanoplatelet stripe waveguides, Micro Nano Eng. 17 (2022) 100167.
	
	\bibitem{bandres2018topological}
	M.~A. Bandres, S.~Wittek, G.~Harari, M.~Parto, J.~Ren, M.~Segev, D.~N.
	Christodoulides, M.~Khajavikhan, Topological insulator laser: Experiments,
	Science 359~(6381) (2018) eaar4005.
	
	\bibitem{el2019dawn}
	R.~El-Ganainy, M.~Khajavikhan, D.~N. Christodoulides, S.~K. Ozdemir, The dawn
	of non-hermitian optics, Commun. Phys. 2~(1) (2019) 37.
	
	\bibitem{ashida2020non}
	Y.~Ashida, Z.~Gong, M.~Ueda, Non-hermitian physics, Adv. Phys. 69~(3) (2020)
	249--435.
	
	\bibitem{zezyulin2011solitonsOL}
	D.~A. Zezyulin, Y.~V. Kartashov, V.~V. Konotop, Solitons in a medium with
	linear dissipation and localized gain, Opt. Lett. 36~(7) (2011) 1200--1202.
	
	\bibitem{wimmer2015observationNC}
	M.~Wimmer, A.~Regensburger, M.-A. Miri, C.~Bersch, D.~N. Christodoulides,
	U.~Peschel, Observation of optical solitons in pt-symmetric lattices, Nat.
	Commun. 6~(1) (2015) 7782.
	
	\bibitem{deconinck2006computing}
	B.~Deconinck, J.~N. Kutz, Computing spectra of linear operators using the
	floquet-fourier-hill method, J. Comput. Phys. 219~(1) (2006) 296--321.
	
	\bibitem{yang2010nonlinear}
	J.~Yang, Nonlinear waves in integrable and nonintegrable systems, SIAM, 2010.
	
	\bibitem{Kang2023}
	J.~Kang, Q.~Zhang, R.~Wei, J.~Qiu, Z.~Yang, G.~Dong, Tunable localization of
	higher-order bound states in non-hermitian optical waveguide lattices, Laser
	Photonics Rev. 17~(12) (2023) 2300558.
	
\end{thebibliography}

\end{document}


\title{Supporting Information for Dissipative surface solitons in two-dimensional truncated lattices with linear gain and loss}

\author{Changming Huang}
\email{hcm123\_2004@126.com}
\affiliation{Department of Physics, Changzhi University, Changzhi, Shanxi 046011, China}
\author{Yan Wang}
\affiliation{Department of Physics, Changzhi University, Changzhi, Shanxi 046011, China}
\author{Pengcheng Liu}
\affiliation{Department of Physics, Changzhi University, Changzhi, Shanxi 046011, China}

\author{Qidong Fu}
\email{jellyfqd@sjtu.edu.cn}
\affiliation{School of Physics and Astronomy, Shanghai Jiao Tong University, Shanghai 200240, China}	
\author{Liangwei Dong}
\email{dlw\_0@163.com}
\affiliation{Department of Physics, Zhejiang University of Science and Technology, Hangzhou, 310023, China}

\date{\today}

\maketitle

\noindent{\normalfont\bfseries S1. \textbf{Derivation of the linear stability eigenvalue problem.}}
\par\vspace{0.25\baselineskip}
We consider the perturbed solution in the form
\begin{equation}
\Psi(x,y,z)=
\left\{
\phi_r(x,y)+i\phi_i(x,y)
+
\left[g_1(x,y)+ig_2(x,y)\right]e^{\tau z}
\right\}e^{i\beta z},
\end{equation}
where $\phi_r$ and $\phi_i$ are the real and imaginary parts of the stationary solution, respectively. The functions $g_1$ and $g_2$ denote the real and imaginary parts of the perturbation amplitude, and $\tau$ is the corresponding perturbation growth rate.

For convenience, we define the stationary field as
\begin{equation}
\phi(x,y)=\phi_r(x,y)+i\phi_i(x,y),
\end{equation}
and introduce the perturbation amplitude
\begin{equation}
\tilde{\eta}(x,y)=g_1(x,y)+ig_2(x,y).
\end{equation}
The full perturbation is then written as
\begin{equation}
\eta(x,y,z)=\tilde{\eta}(x,y)e^{\tau z}.
\end{equation}
Therefore, the perturbed field can be expressed compactly as
\begin{equation}
\Psi(x,y,z)=\left[\phi(x,y)+\eta(x,y,z)\right]e^{i\beta z}
=
\left[\phi(x,y)+\tilde{\eta}(x,y)e^{\tau z}\right]e^{i\beta z}.
\end{equation}

The governing equation is
\begin{equation}
i\frac{\partial \Psi}{\partial z}
=
-\frac{1}{2}
\left(
\frac{\partial^2\Psi}{\partial x^2}
+
\frac{\partial^2\Psi}{\partial y^2}
\right)
-\left(V_r-iV_i\right)\Psi
-\sigma |\Psi|^2\Psi .
\end{equation}
Introducing the two-dimensional Laplacian operator
\begin{equation}
\Delta=
\frac{\partial^2}{\partial x^2}
+
\frac{\partial^2}{\partial y^2},
\end{equation}
the governing equation can be written as
\begin{equation}
i\frac{\partial \Psi}{\partial z}
=
-\frac{1}{2}\Delta\Psi
-\left(V_r-iV_i\right)\Psi
-\sigma |\Psi|^2\Psi .
\end{equation}

Substituting
\begin{equation}
\Psi=\left(\phi+\eta\right)e^{i\beta z}
\end{equation}
into the governing equation gives
\begin{equation}
i
\left[
\frac{\partial \eta}{\partial z}
+i\beta(\phi+\eta)
\right]
=
-\frac{1}{2}\Delta(\phi+\eta)
-\left(V_r-iV_i\right)(\phi+\eta)
-\sigma|\phi+\eta|^2(\phi+\eta).
\end{equation}
Since
\begin{equation}
\eta(x,y,z)=\tilde{\eta}(x,y)e^{\tau z}, \nonumber
\end{equation}
we have
\begin{equation}
\frac{\partial \eta}{\partial z}
=
\tau \tilde{\eta}e^{\tau z}
=
\tau\eta .
\end{equation}
Thus,
\begin{equation}
i\tau\eta-\beta(\phi+\eta)
=
-\frac{1}{2}\Delta(\phi+\eta)
-\left(V_r-iV_i\right)(\phi+\eta)
-\sigma|\phi+\eta|^2(\phi+\eta).
\end{equation}

The stationary solution $\phi$ satisfies
\begin{equation}
-\beta\phi
=
-\frac{1}{2}\Delta\phi
-\left(V_r-iV_i\right)\phi
-\sigma|\phi|^2\phi .
\end{equation}
Subtracting the stationary equation from the perturbed equation yields
\begin{equation}
i\tau\eta-\beta\eta
=
-\frac{1}{2}\Delta\eta
-\left(V_r-iV_i\right)\eta
-\sigma\left[
|\phi+\eta|^2(\phi+\eta)-|\phi|^2\phi
\right].
\end{equation}

Next, we linearize the nonlinear term with respect to the perturbation $\eta$. To first order in $\eta$, one has
\begin{equation}
|\phi+\eta|^2
=
(\phi+\eta)(\phi^*+\eta^*)
=
|\phi|^2+\phi^*\eta+\phi\eta^*
+O(|\eta|^2).
\end{equation}
Therefore,
\begin{align}
|\phi+\eta|^2(\phi+\eta)
&=
\left(
|\phi|^2+\phi^*\eta+\phi\eta^*
\right)(\phi+\eta)
+O(|\eta|^2) \notag\\
&=
|\phi|^2\phi
+
|\phi|^2\eta
+
\phi^*\eta\phi
+
\phi\eta^*\phi
+
O(|\eta|^2) \notag\\
&=
|\phi|^2\phi
+
2|\phi|^2\eta
+
\phi^2\eta^*
+
O(|\eta|^2).
\end{align}
Thus, after retaining only the linear terms, we obtain
\begin{equation}
|\phi+\eta|^2(\phi+\eta)-|\phi|^2\phi
=
2|\phi|^2\eta+\phi^2\eta^*.
\end{equation}
The linearized perturbation equation is therefore
\begin{equation}
i\tau\eta-\beta\eta
=
-\frac{1}{2}\Delta\eta
-\left(V_r-iV_i\right)\eta
-\sigma
\left(
2|\phi|^2\eta+\phi^2\eta^*
\right).
\end{equation}
Equivalently,
\begin{equation}
i\tau\eta
=
\left[
-\frac{1}{2}\Delta
-V_r+iV_i+\beta
-2\sigma|\phi|^2
\right]\eta
-\sigma\phi^2\eta^* .
\end{equation}

Since
\begin{equation}
\eta=\tilde{\eta}e^{\tau z},
\qquad
\eta^*=\tilde{\eta}^*e^{\tau z},
\end{equation}
and the linearized equation is homogeneous in the perturbation, every term contains the common factor $e^{\tau z}$. Dividing out this factor gives the equation for the perturbation amplitude:
\begin{equation}
i\tau\tilde{\eta}
=
\left[
-\frac{1}{2}\Delta
-V_r+iV_i+\beta
-2\sigma|\phi|^2
\right]\tilde{\eta}
-\sigma\phi^2\tilde{\eta}^* .
\end{equation}

We now decompose this equation into its real and imaginary parts. By definition,
\begin{equation}
\phi=\phi_r+i\phi_i,
\qquad
\tilde{\eta}=g_1+ig_2.
\end{equation}
Thus,
\begin{equation}
|\phi|^2=\phi_r^2+\phi_i^2,
\end{equation}
and
\begin{equation}
\phi^2
=
(\phi_r+i\phi_i)^2
=
\phi_r^2-\phi_i^2+2i\phi_r\phi_i .
\end{equation}
Moreover,
\begin{equation}
\tilde{\eta}^*=g_1-ig_2.
\end{equation}
Therefore,
\begin{align}
\phi^2\tilde{\eta}^*
&=
\left(
\phi_r^2-\phi_i^2+2i\phi_r\phi_i
\right)
\left(
g_1-ig_2
\right)=
\left[
(\phi_r^2-\phi_i^2)g_1
+
2\phi_r\phi_i g_2
\right]
+i
\left[
2\phi_r\phi_i g_1
-
(\phi_r^2-\phi_i^2)g_2
\right].
\end{align}

The left-hand side of the perturbation-amplitude equation becomes
\begin{equation}
i\tau\tilde{\eta}
=
i\tau(g_1+ig_2)
=
-\tau g_2+i\tau g_1 .
\end{equation}
Hence, its real and imaginary parts are
\begin{equation}
\operatorname{Re}(i\tau\tilde{\eta})=-\tau g_2,
\qquad
\operatorname{Im}(i\tau\tilde{\eta})=\tau g_1.
\end{equation}

The first term on the right-hand side is
\begin{equation}
\left[
-\frac{1}{2}\Delta
-V_r+\beta
-2\sigma(\phi_r^2+\phi_i^2)
+iV_i
\right](g_1+ig_2).
\end{equation}
Its real part is
\begin{equation}
\left[
-\frac{1}{2}\Delta
-V_r+\beta
-2\sigma(\phi_r^2+\phi_i^2)
\right]g_1
-
V_i g_2,
\end{equation}
and its imaginary part is
\begin{equation}
\left[
-\frac{1}{2}\Delta
-V_r+\beta
-2\sigma(\phi_r^2+\phi_i^2)
\right]g_2
+
V_i g_1.
\end{equation}

The second term on the right-hand side is
\begin{equation}
-\sigma\phi^2\tilde{\eta}^*.
\end{equation}
According to the expansion above, its real part is
\begin{equation}
-\sigma
\left[
(\phi_r^2-\phi_i^2)g_1
+
2\phi_r\phi_i g_2
\right],
\end{equation}
and its imaginary part is
\begin{equation}
-\sigma
\left[
2\phi_r\phi_i g_1
-
(\phi_r^2-\phi_i^2)g_2
\right].
\end{equation}

Combining the real parts gives
\begin{align}
-\tau g_2
&=
\left[
-\frac{1}{2}\Delta
-V_r+\beta
-2\sigma(\phi_r^2+\phi_i^2)
\right]g_1
-
V_i g_2 
-\sigma
\left[
(\phi_r^2-\phi_i^2)g_1
+
2\phi_r\phi_i g_2
\right].
\end{align}
After collecting the coefficients of $g_1$ and $g_2$, we obtain
\begin{equation}
-\tau g_2
=
\left[
-\frac{1}{2}\Delta
-V_r+\beta
-\sigma(3\phi_r^2+\phi_i^2)
\right]g_1
-
\left(
V_i+2\sigma\phi_r\phi_i
\right)g_2 .
\end{equation}
Multiplying both sides by $-1$ yields
\begin{equation}
\tau g_2
=
\left[
\frac{1}{2}\Delta
+V_r-\beta
+\sigma(3\phi_r^2+\phi_i^2)
\right]g_1
+
\left(
V_i+2\sigma\phi_r\phi_i
\right)g_2 .
\end{equation}

Similarly, combining the imaginary parts gives
\begin{align}
\tau g_1
&=
\left[
-\frac{1}{2}\Delta
-V_r+\beta
-2\sigma(\phi_r^2+\phi_i^2)
\right]g_2
+
V_i g_1
-\sigma
\left[
2\phi_r\phi_i g_1
-
(\phi_r^2-\phi_i^2)g_2
\right].
\end{align}
Collecting the coefficients of $g_1$ and $g_2$, we obtain
\begin{equation}
\tau g_1
=
\left[
-\frac{1}{2}\Delta
-V_r+\beta
-\sigma(\phi_r^2+3\phi_i^2)
\right]g_2
+
\left(
V_i-2\sigma\phi_r\phi_i
\right)g_1 .
\end{equation}
Equivalently,
\begin{equation}
\tau g_1
=
\left[
-\frac{1}{2}\Delta
-V_r+\beta
-\sigma(\phi_r^2+3\phi_i^2)
\right]g_2
-
\left(
2\sigma\phi_r\phi_i-V_i
\right)g_1 .
\end{equation}

Finally, replacing $\Delta$ by its explicit expression, the linear stability eigenvalue problem can be written as
\begin{align}
\tau g_1
&=
\left[
-\frac{1}{2}
\left(
\frac{\partial^2}{\partial x^2}
+
\frac{\partial^2}{\partial y^2}
\right)
-V_r+\beta
-\sigma\left(\phi_r^2+3\phi_i^2\right)
\right]g_2
-
\left(
2\sigma\phi_r\phi_i-V_i
\right)g_1,
\\
\tau g_2
&=
\left[
+\frac{1}{2}
\left(
\frac{\partial^2}{\partial x^2}
+
\frac{\partial^2}{\partial y^2}
\right)
+V_r-\beta
+\sigma\left(3\phi_r^2+\phi_i^2\right)
\right]g_1
+
\left(
2\sigma\phi_r\phi_i+V_i
\right)g_2 .
\end{align}

\par\vspace{0.25\baselineskip}
\noindent{\normalfont\bfseries S2. \textbf{Linear surface modes at the interface.}}
\par\vspace{0.25\baselineskip}

In the linear regime, the functions of linear modes for $\sigma = 0$ of Eq.~(1) are sought as $\Psi(x,y,z)=\phi(x,y)e^{i\beta_\text{lin} z}$, here, $\phi(x,y)$ is the linear mode, and $\beta_\text{lin}$ is its corresponding eigenvalue (propagation constant). Substituting this expression into Eq. (1), one can get
\begin{align}\label{eqs1}
\beta_\text{lin}\phi(x,y) =\mathcal{H}\phi(x,y);\quad \mathcal{H}= \frac{1}{2}\left(\frac{\partial^2}{\partial x^2}+\frac{\partial^2}{\partial y^2}\right)+(V_r-iV_i).
\end{align}
This linear eigenvalue problem (\ref{eqs1}) can be solved numerically using a standard eigenvalue solver.

\begin{figure}[htbp]
\centering
\includegraphics[width=1\columnwidth]{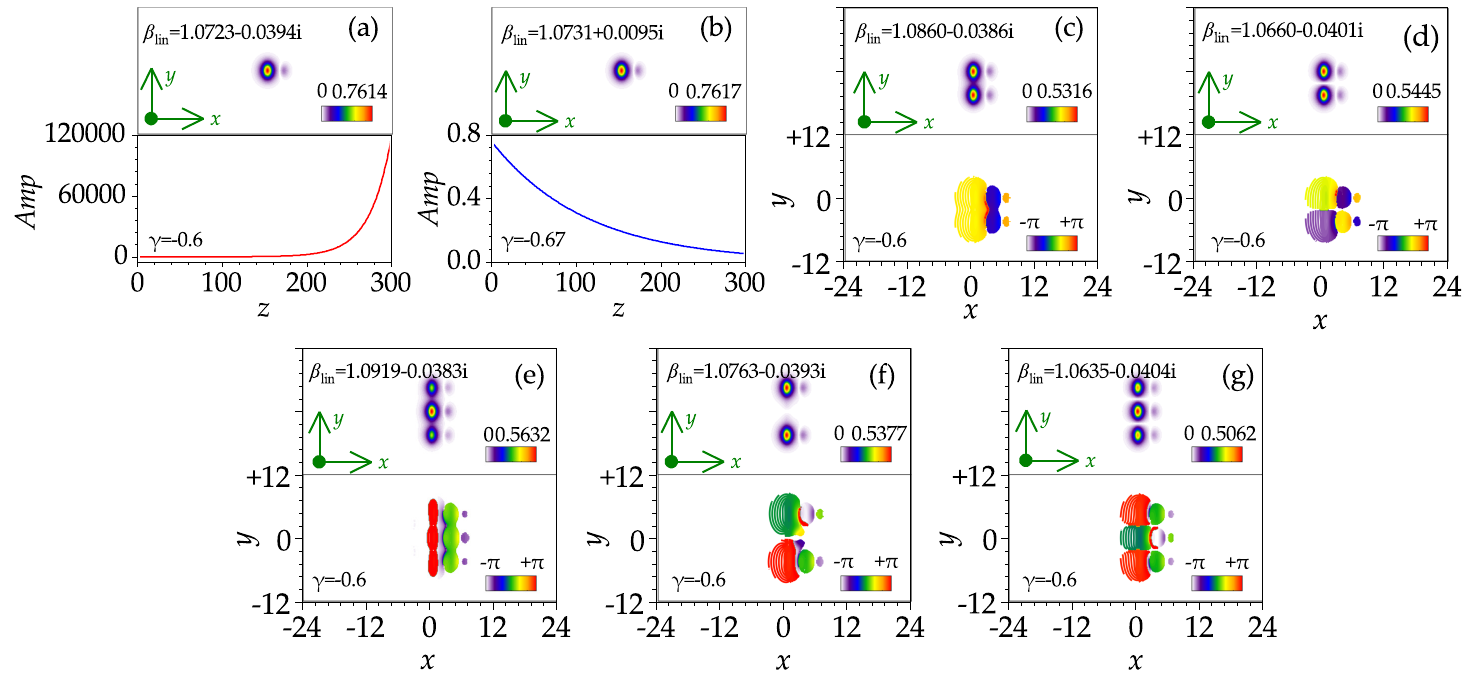}
\caption{
Fundamental gain (a) and loss (b) mode profiles (upper panels) and their propagation dynamics (lower panels) in a transversely truncated single-waveguide array. Localized gap modes in dual- (c,d) and triple-truncated (e–g) waveguide arrays; upper panels show field modulus distributions and lower panels show phase profiles.
}
\label{figSI1}
\end{figure}

To elucidate the optical characteristics of surface dissipative solitons, it is crucial to examine the modal profiles and eigenvalue spectra of the associated surface modes.
In a transversely truncated single-waveguide array, a fundamental localized surface mode is confined at the interface for a given parameter $\gamma$. The real and imaginary parts of its eigenvalue both decrease monotonically as $\gamma$ increases [Fig.~\textcolor{blue}{1}(d) in the main text]. A critical value $\gamma_0$ delineates two distinct regimes: for $\gamma < \gamma_0$, the mode exhibits net loss, whereas for $\gamma > \gamma_0$, it becomes gain-dominated. 
Representative gain and loss field profiles of the fundamental mode ($\phi(x,y)$ is normalized as $\int_{-\infty}^{+\infty}\int_{-\infty}^{+\infty}|\phi(x,y)|^2dxdy=1$) are shown in the upper panels of Figs.~\ref{figSI1}(a,b), with their corresponding dynamical evolutions presented in the lower panels.
By considering the factor $e^{i\beta_\text{lin} z} = e^{i(\beta_\text{lin,r}+i\beta_\text{lin,i})z}=e^{i\beta_\text{lin,r}z}\cdot e^{-\beta_\text{lin,i} z}$, it can also be inferred that $\beta_\text{lin,i}<0$ corresponds to a gain mode, whereas $\beta_\text{lin,i}>0$ represents a loss mode.

In the transversely truncated two-waveguide array, two distinct families of surface modes emerge at the interface: an in-phase mode and an out-of-phase mode. These modes exhibit markedly different field profiles and phase structures, as illustrated in Figs.~\ref{figSI1}(c,d).
The modal profiles show that the in-phase mode retains a finite field amplitude at $x=0$ and $y=-d/2$, while the out-of-phase mode develops a node at this point. The phase contour maps confirm these symmetries uniform color regions indicate in-phase behavior, whereas alternating color patterns signify out-of-phase characteristics.
As the number of waveguides truncated along the $y$-axis increases, the number of supported surface modes correspondingly grows. In a transversely truncated three-waveguide array, in addition to the in-phase and out-of-phase modes, an antisymmetric mode also emerges [Fig.~\ref{figSI1}(e-g)]. In this dissipative system, the eigenvalue of the in-phase mode is generally slightly higher than that of the out-of-phase mode, while the newly emerging surface mode possesses an eigenvalue that resides between them [Figs.~\textcolor{blue}{1}(e,f) in the main text]. The antisymmetric mode features intensity peaks localized in the two outermost waveguides. These enriched modal structures provide valuable guidance for understanding the behavior of nonlinear dissipative surface solitons.

\par\vspace{0.25\baselineskip}
\noindent{\normalfont\bfseries S3. \textbf{Propagation examples of fundamental DSSs in a single truncated array.}}
\par\vspace{0.25\baselineskip}

\begin{figure*}[htbp]
	\centering
	\includegraphics[width=0.75\columnwidth]{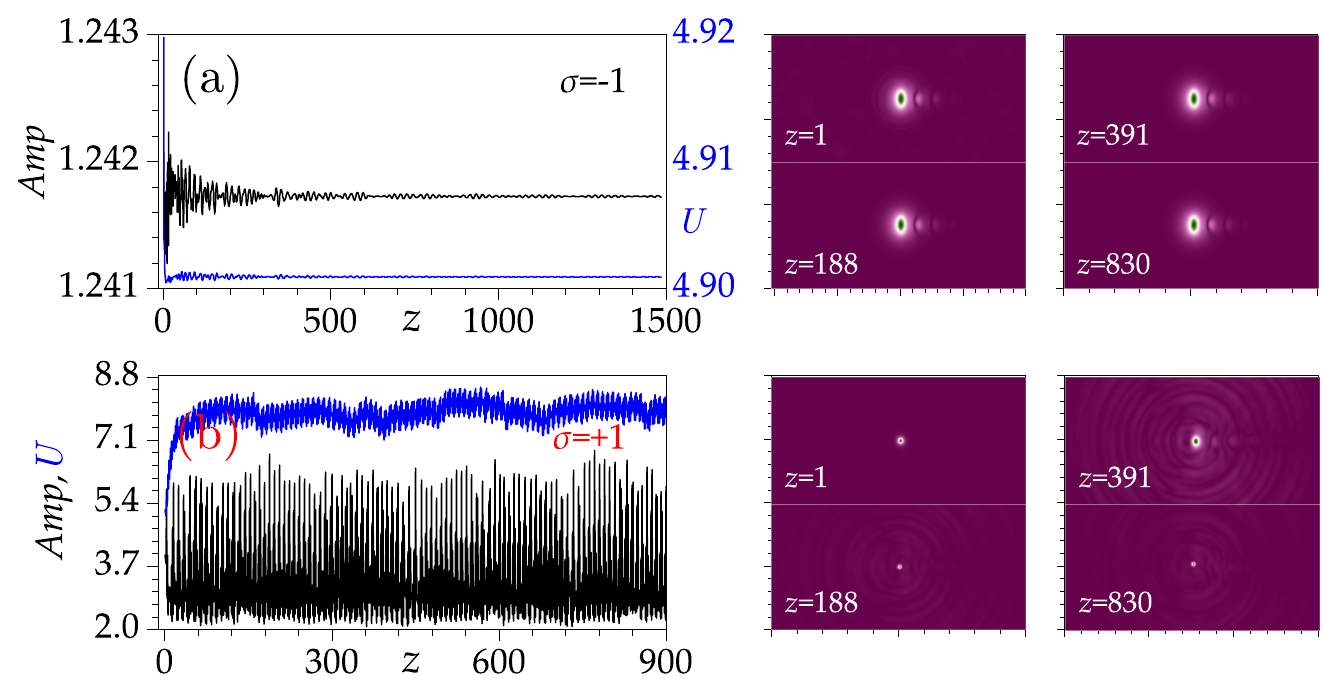}
	\caption{Propagation dynamics of fundamental dissipative solitons under defocusing (a) and focusing (b) nonlinearities. The evolution of the soliton amplitude $Amp$ (black curves) and power $U$ (blue curves) as functions of the propagation distance $z$ is shown, and the right panels display the corresponding field modulus profiles at different propagation distances $z$. $\gamma=-0.6$ in (a) and (b).
}
	\label{figSI2}
\end{figure*}

Figure \ref{figSI2} illustrates the propagation dynamics of fundamental dissipative surface solitons under defocusing ($\sigma=-1$) and focusing ($\sigma=+1$) nonlinearities. A random perturbation with an amplitude of $10\%$ is added to the input beam.
In panel (a), corresponding to the defocusing case, the evolution of the soliton amplitude $Amp$ (black curve) and power $U$ (blue curve) as functions of the propagation distance $z$ is shown. After a short distance characterized by weak oscillations, both quantities rapidly converge to constant values and remain nearly unchanged over long propagation distances. The accompanying field modulus snapshots at representative propagation distances ($z=1$, $188$, $391$, and $830$) further confirm that the field profile preserves its shape and localization, demonstrating stable propagation supported by the combined action of the surface lattice, linear gain and loss, and defocusing nonlinearity. In contrast, panel (b) displays the propagation behavior in the focusing regime. Here, both the amplitude and the power exhibit pronounced oscillations throughout propagation, without settling into a steady-state value. The corresponding field distributions reveal the distortion and the emergence of radiative wave patterns, reflecting strong breathing dynamics and power redistribution. These features indicate that the fundamental dissipative soliton becomes dynamically unstable under focusing nonlinearity.

\par\vspace{0.25\baselineskip}
\noindent{\normalfont\bfseries S4. \textbf{Phase, phase-gradient, and power-flow of IP and OOP DSSs in dual truncated arrays.}}
\par
\begin{figure*}[htbp]
	\centering
	\includegraphics[width=0.75\columnwidth]{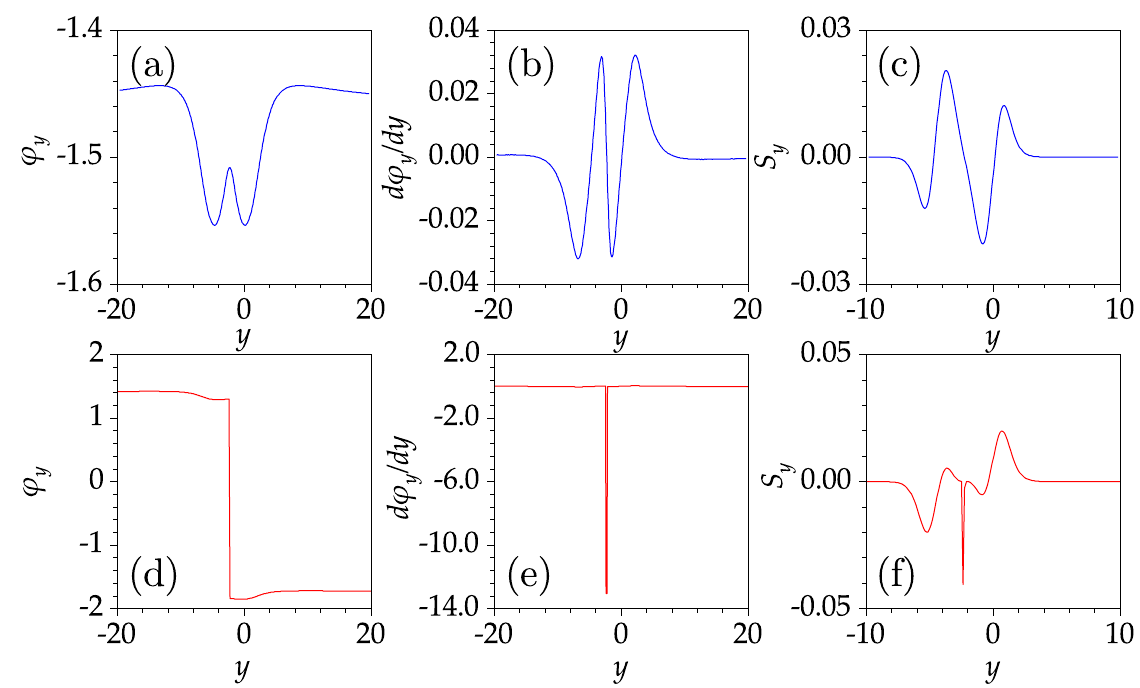}
	\caption{Phase $\varphi_y$ (a,d), transverse phase-gradient $d\varphi_y/dy$ (b,e), and power-flow $S_y$ (c,f) distributions of the in-phase (a-c) and out-of-phase (d-f) dissipative surface solitons along $y$-direction. $\gamma=-0.6$ and $\sigma=-1$ in all panels.
}
	\label{figSI3}
\end{figure*}

\vspace{0.25\baselineskip}
For both in-phase and out-of-phase dissipative surface solitons, we examined their $y$-direction profiles at $x=0$. The corresponding phase $\varphi_y=\arctan[\phi_i(x=0,y),\phi_r(x=0,y)]$, phase-gradient $d\varphi_y/dy$, and power-flow $S_y=|\phi(x=0,y)|^2d\varphi_y/dy$ distributions are shown in Fig.~\ref{figSI3}. For in-phase dissipative surface solitons, the $y$-direction phase remains nearly smooth, showing no discernible jump between the two lobes [Fig.~\ref{figSI3}(a)]. The transverse phase-gradient is correspondingly weak, with only slight oscillations near the peaks [Fig.~\ref{figSI3}(b)], resulting in merely small localized variations in the power-flow and essentially no net transverse power transfer [Fig.~\ref{figSI3}(c)]. These features allow the in-phase surface soliton to retain a stable steady state under defocusing nonlinearity and the surface lattice. In contrast, out-of-phase dissipative surface solitons exhibit a pronounced phase step--close to $\pi$--between the lobes [Fig.~\ref{figSI3}(d)]. This sharp discontinuity generates a strongly localized phase slope near $y=-d/2$ [Fig.~\ref{figSI3}(e)], which produces distinct positive and negative power-flow peaks in that narrow region. The associated intense transverse power-flow enhance delocalization in the presence of defocusing nonlinearity, and together with the gain (or loss) array configuration along the $y$ direction, ultimately driving out-of-phase surface soliton unstable [Fig.~\ref{figSI3}(f)].

\begin{figure*}[htbp]
	\centering
	\includegraphics[width=0.75\columnwidth]{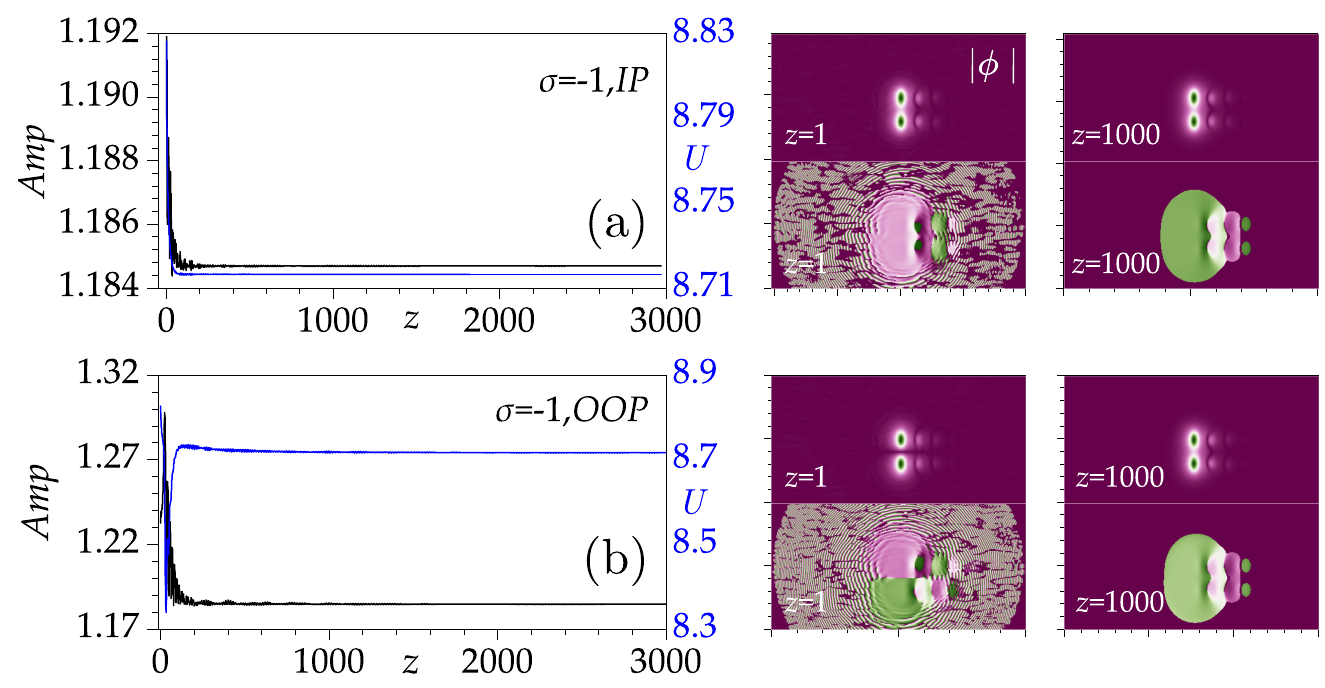}
	\caption{Propagation dynamics of in-phase (a) and out-of-phase (b) dissipative solitons. The evolution of the soliton amplitude $Amp$ (black curves) and power $U$ (blue curves) with the propagation distance $z$, and the right panels display the corresponding field modulus $|\phi|$ and phase $\varphi$ at different propagation distances $z$. $\gamma=-0.6$ and $\sigma=-1$ in (a) and (b).
}
	\label{figSI4}
\end{figure*}


\par\vspace{0.25\baselineskip}
\noindent{\normalfont\bfseries S5. \textbf{Propagation examples of IP and OOP DSSs in dual truncated arrays.}}
\par\vspace{0.25\baselineskip}
The propagation examples of in-phase and out-of-phase dissipative surface solitons for $\sigma=-1$ are shown in Fig.~\ref{figSI4}. Panels (a) and (b) correspond to the in-phase and out-of-phase configurations, respectively. In each case, the evolution of peak amplitude $Amp$ (black curves) and power $U$ (blue curves) with propagation distance $z$ is depicted.

For the in-phase dissipative surface soliton [panel (a)], both the amplitude and power exhibit only weak transient oscillations at the initial stage and rapidly relax to stationary values. They remain essentially unchanged over long propagation distances up to $z=3000$, indicating a robust behavior of stable dissipative solitons. The accompanying field-modulus snapshots $|\phi|$ and phase snapshots $\varphi$ at representative propagation distances ($z=1$ and $z=1000$) confirm that the soliton preserves its localized structure and symmetry. 

In contrast, for the out-of-phase dissipative surface soliton [panel (b)], the amplitude and power rapidly stabilize to a steady-state value following an initial transient regime.
The phase distributions initially exhibit an out-of-phase configuration, characterized by a phase discontinuity between the two lobes. As the amplitude and power converge to constant values, the phase profile gradually transitions to an in-phase distribution. This behavior suggests that the initially unstable out-of-phase dissipative surface solitons evolve into stable in-phase dissipative surface solitons.


\par\vspace{0.25\baselineskip}
\noindent{\normalfont\bfseries S6. \textbf{Instability dynamics near the stability boundaries in different truncated waveguide arrays.}}
\par
\begin{figure*}[htbp]
	\centering
	\includegraphics[width=0.75\columnwidth]{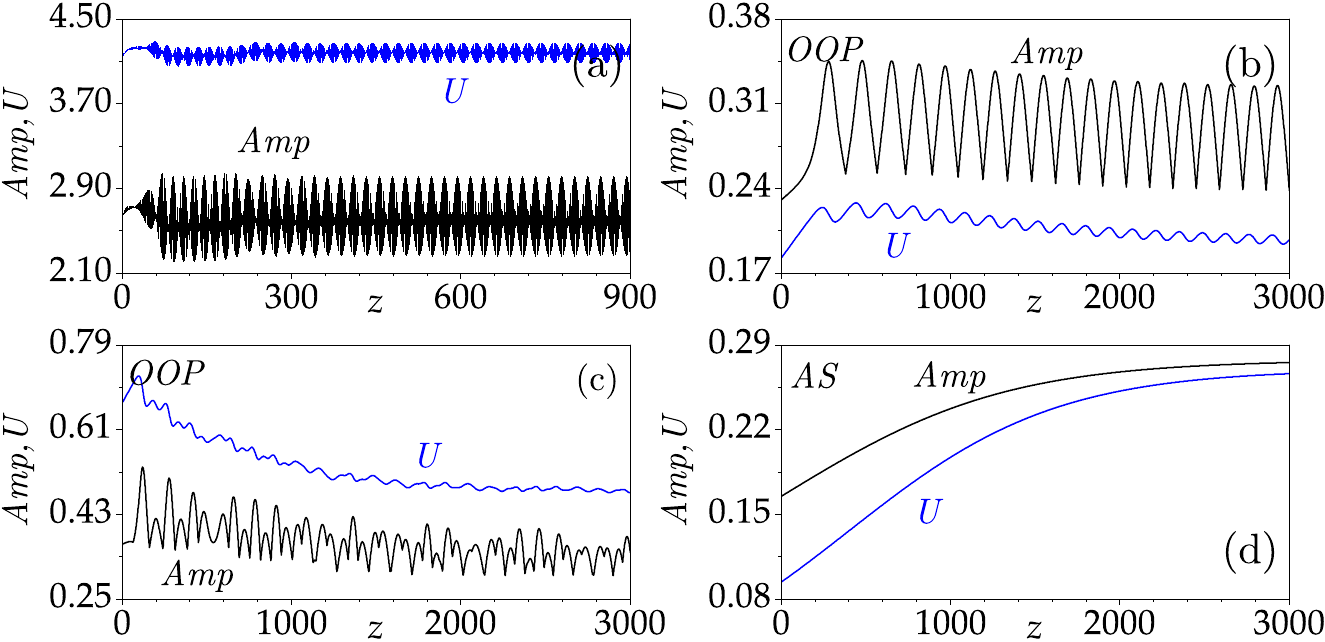}
	\caption{Propagation dynamics of unstable DSSs in different truncated complex lattices. Black and blue curves show the peak amplitude $Amp$ and power $U$, respectively. (a) Fundamental surface soliton in a single truncated array, $\sigma=+1$, and $\gamma=-0.646$; (b) OOP DSS in a dual truncated array, $\gamma=-0.656$; (c) OOP DSS in a triple truncated array, $\gamma=-0.655$; and (d) AS in a triple truncated array, $\gamma=-0.656$. $\sigma=-1$ in (b-d).
}
	\label{figSI5}
\end{figure*}

\vspace{0.25\baselineskip}
The propagation of representative unstable states near the stability boundaries is shown in Fig.~\ref{figSI5}. For the fundamental soliton in the single truncated array, both the peak amplitude $Amp$ and the power $U$ undergo pronounced oscillations during propagation [Fig.~\ref{figSI5}(a)]. The state remains spatially localized over the simulated distance, but it does not relax to a stationary attractor. Instead, the instability appears as a breathing-type dynamics, indicating a weak oscillatory instability close to the stability edge.

A different behaviour is found for the out-of-phase states in the dual and triple truncated arrays. As shown in Figs.~\ref{figSI5}(b,c), the peak amplitude oscillates persistently, whereas the total power decreases in an oscillatory fashion. 
During propagation, the localized peaks residing on different rows remain coupled and continuously exchange power. After sufficiently long propagation, this inter-row coupling drives the field towards a phase-locked configuration: both the amplitude and the power gradually approach well-defined stationary values, and the initially out-of-phase state eventually relaxes into a stable in-phase dissipative surface soliton.

For the asymmetric state in the triple truncated array, the instability follows another route [Fig.~\ref{figSI5}(d)]. Both the amplitude $Amp$ and the power $U$ increase slowly with propagation distance $z$. Nevertheless, the corresponding field profile $|\phi|$ remains well localized over the finite propagation distance considered here, namely $z\leq 3000$. This behaviour is consistent with the fact that the selected state lies close to the stability boundary, where the associated instability growth rate is small. As a result, the unstable perturbation develops only gradually, allowing the soliton to preserve its localized structure over a long but finite propagation distance.

\par\vspace{0.25\baselineskip}
\noindent{\normalfont\bfseries S7. \textbf{Weak-input excitation of stable dissipative surface solitons.}}
\par
\begin{figure*}[htbp]
	\centering
	\includegraphics[width=0.75\columnwidth]{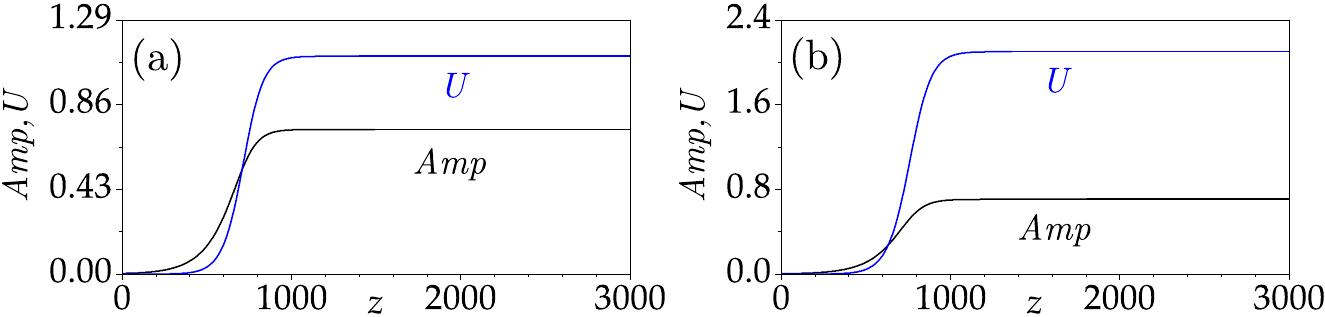}
	\caption{Evolution of the peak amplitude $Amp$ (Black curves) and power $U$ (Blue curves) under weak input excitation in truncated complex waveguide arrays. (a) in a single truncated array, and (b) in a dual truncated array. $A=0.01,p_r=4,p_i=1,d=4.5,w_0=0.5,\gamma=-0.646,\sigma=-1$ in (a) and (b).
}
	\label{figSI6}
\end{figure*}

\vspace{0.25\baselineskip}
To clarify whether the excitation of stable dissipative surface solitons has a rigid threshold character, we performed additional propagation simulations using weak Gaussian inputs close to the zero-intensity background. The results are shown in Fig.~\ref{figSI6}. In the single truncated array, a low-amplitude Gaussian beam $\Psi(x,y,z=0)=A\exp[-(x^2+y^2)/w_0^2]$ with $A=0.01,w_0=0.5$ is gradually amplified during propagation and evolves into a stable fundamental dissipative surface soliton [Fig.~\ref{figSI6}(a)]. Both the peak amplitude $Amp$ and the power $U$ increase from nearly zero and then saturate to well-defined stationary values.

A similar excitation process is observed in the dual truncated array. When a weak in-phase Gaussian beam array $\Psi(x,y,z=0)=A\exp[-(x^2+y^2)/w_0^2]+A\exp[-(x^2+(y+d)^2)/w_0^2]$ with the same $A$ and $w_0$ is launched into the structure, the field evolves into a stable in-phase dissipative surface soliton [Fig.~\ref{figSI6}(b)]. Again, both $Amp$ and $U$ grow from the weak input level and eventually approach constant values, indicating the formation of a stationary nonlinear dissipative state.

These results show that, in the defocusing regime considered here, the excitation of stable dissipative surface solitons is not strictly rigid or threshold-like. Instead, the stable states can be dynamically reached from weak perturbations close to the zero background. The weak input is first amplified by the localized gain, then reshaped by diffraction, lattice confinement and Kerr self-action, and finally arrested by the global gain-loss balance.

